\documentclass[fleqn,usenatbib]{mnras}

\usepackage[T1]{fontenc}
\usepackage{float}

\DeclareRobustCommand{\VAN}[3]{#2}
\let\VANthebibliography\thebibliography
\def\thebibliography{\DeclareRobustCommand{\VAN}[3]{##3}\VANthebibliography}

\usepackage{graphicx}	% Including figure files
\usepackage{amsmath}	% Advanced maths commands
\usepackage{amssymb}
\usepackage{array}
\usepackage{verbatim}
\usepackage{makecell}
\usepackage{orcidlink}
\usepackage{tabularx}
\usepackage{hyperref}% add hypertext capabilities

\title[LSB Galaxies at 0.4 < z < 0.8 in GOODS-S]{JADES: Low Surface Brightness Galaxies at 0.4 < z < 0.8 in GOODS-S}

\author[Shields et al.]{Tristen Shields\orcidlink{0009-0000-5273-7870},$^{1}$\thanks{E-mail: tdshield@arizona.edu}
Marcia Rieke\orcidlink{0000-0002-7893-6170},$^{1}$\thanks{E-mail: mrieke@gmail.com}
Kevin Hainline\orcidlink{0000-0003-4565-8239},$^{1}$
Jakob M. Helton\orcidlink{0000-0003-4337-6211},$^{2}$
\newauthor Andrew J.\ Bunker\orcidlink{0000-0002-8651-9879},$^{3}$
Courtney Carreira\orcidlink{0000-0001-6301-3667},$^{4}$
Emma Curtis-Lake\orcidlink{0000-0002-9551-0534},$^{5}$
Daniel J.\ Eisenstein\orcidlink{0000-0002-2929-3121},$^{6}$
\newauthor Benjamin D.\ Johnson\orcidlink{0000-0002-9280-7594},$^{6}$
Pierluigi Rinaldi\orcidlink{0000-0002-5104-8245},$^{7}$
Brant Robertson\orcidlink{0000-0002-4271-0364},$^{4}$
Christina C. Williams\orcidlink{0000-0003-2919-7495},$^{8}$
\newauthor Christopher N. A. Willmer\orcidlink{0000-0001-9262-9997},$^{1}$
Yang Sun\orcidlink{0000-0001-6561-9443}$^{1}$
\\
$^{1}$Steward Observatory, University of Arizona, 933 N Cherry Ave, Tucson, AZ, 85721, USA\\
$^{2}$Department of Astronomy \& Astrophysics, The Pennsylvania State University, University Park, PA 16802, USA\\
$^{3}$Department of Physics, University of Oxford, Denys Wilkinson Building, Keble Road, Oxford OX1 3RH, UK\\
$^{4}$Department of Astronomy and Astrophysics, University of California, Santa Cruz, 1156 High Street, Santa Cruz, CA 95064, USA\\
$^{5}$Centre for Astrophysics Research, Department of Physics, Astronomy and Mathematics, University of Hertfordshire, Hatfield AL10 \\9AB, UK\\
$^{6}$Center for Astrophysics $|$ Harvard \& Smithsonian, 60 Garden St., Cambridge MA 02138 USA\\
$^{7}$Space Telescope Science Institute, 3700 San Martin Drive, Baltimore, Maryland 21218, USA\\
$^{8}$NSF National Optical-Infrared Astronomy Research Laboratory, 950 North Cherry Avenue, Tucson, AZ 85719, USA\\
}

\date{Accepted XXX. Received YYY; in original form ZZZ}

\pubyear{2025}

\begin{document}
\label{firstpage}
\pagerange{\pageref{firstpage}--\pageref{lastpage}}
\maketitle

\begin{abstract}
Low surface brightness galaxies (LSBs) are an important class of galaxies that allow us to broaden our understanding of galaxy formation and test various cosmological models. We present a survey of low surface brightness galaxies at $0.4 < z_{\rm phot} <  0.8$ in the GOODS-S field using JADES data. We model LSB surface brightness profiles, identifying those with $\bar{\mu}_{\rm eff} > 24$ mag arcsec$^{-2}$ in the F200W JWST/NIRCam filter. We study the spatial distribution, number density, S\'{e}rsic profile parameters, and rest-frame colours of these LSBs. We compare the photometrically-derived star formation histories, mass-weighted ages, and dust attenuations of these galaxies with a high surface brightness (HSB) sample at similar redshift and a lower redshift ($z_{\rm phot} < 0.4$) LSB sample, all of which have stellar masses $\lesssim 10^8 M_{\odot}$. We find that all samples have low star formation (SFR$_{100} \lesssim 0.01 \: M_{\odot} \: \textup{yr}^{-1}$). The higher redshift LSBs and HSBs have similar star formation histories which show that the LSBs and HSBs possibly come from the same progenitors at $z \gtrsim 2$, though the histories are not well constrained for the LSB samples. The LSBs appear to have minimal dust, with most of our LSB samples showing $A_V < 1$ mag. JWST has pushed our understanding of LSBs beyond the local Universe.
\end{abstract}


%%%%%%%%%%%%%%%%%%%%%%%%%%%%%%%%%%%%%%%%%%%%%%%%%%

%%%%%%%%%%%%%%%%% BODY OF PAPER %%%%%%%%%%%%%%%%%%

\section{Introduction}
\label{sec:intro}
Low surface brightness galaxies (LSBs) are galaxies with a surface brightness dimmer than that of the ground-based night sky, and as a result have been historically difficult to detect and observe. Early research on LSBs suggested a lower limit to a galaxy's surface brightness \citep{freeman1970}. It was then suggested that this was not a physical law, but rather an effect of selection bias due to detection limits at the time \citep{disney1976}. Since then, it has been confirmed that galaxies dimmer than the ground-based sky background exist, and constitute a sizeable fraction (up to 60\%) of the local galaxy population \citep{mcgaugh1996, oneil_bothun2000, minchin2004}. We present here a search for LSBs at moderate ($0.4 < z < 0.8$) redshift using JWST. We describe how these newly observed objects fit into the context of older LSB definitions.

The quantitative definition of a ``low surface brightness'' galaxy varies from study to study and is primarily based on the detection limits of a given survey. When looking for LSBs, the goal is to find relatively faint and somewhat extended objects. In the past, LSBs have typically been defined as having a $B$-band central surface brightness of $\mu_0(B) \ge 23$ mag arcsec$^{-2}$ \citep{bothun1997, impey_bothun1997}. More recent studies have used mean effective surface brightness to consider high surface brightness (HSB) bulges in LSB disks, as relying on central surface brightness can lead to exclusion of such disks \citep{greco_etal2018}. 

Common characteristics of LSBs from past surveys indicate that LSBs are diverse in color, morphology, and environment. LSBs in past surveys have low star formation rates (SFRs) \citep{van_den_hoek2000, burkholder2001}, are abundant in neutral hydrogen (HI) \citep{schombert1992, o'neil2004} but deficient in ionized hydrogen (HII) \citep{mcgaugh1995}, are metal poor \citep{mcgaugh1994, de_blok1998}, dust poor \citep{rahman2007}, and lack molecular gas (CO) \citep{de_blok1998b}. Defining LSBs as having a mean effective $g$-band surface brightness of $\bar{\mu}_{\textup{eff}}(g)$ > 24.3 mag arcsec$^{-2}$, \cite{greco_etal2018} studied $\sim$700 LSB galaxies in the local Universe with the Hyper Suprime-Cam Subaru Strategic programme (HSC-SSP). They effectively split their sample into blue and red populations. While these LSBs spanned many different environments and morphologies, the blue LSBs in their sample had irregular morphologies with signs of ongoing star formation, while the red LSBs tended to have smooth light profiles well-characterized by a single S\'{e}rsic profile, with signs of older stellar populations.

The photometric spectral energy distributions (SEDs) of LSBs have been well fit by old (> 7 Gyr) stellar populations \citep{jimenez1998} which have been suggested to be the primary source of dust heating in LSBs \citep{rahman2007}. LSBs also form the majority of galaxies in the dwarf ($M_* < 10^9 M_{\odot}$) regime \citep{jackson2021}. A recent analysis of the Horizon AGN hydrodynamical simulation \citep{dubois2014, kaviraj2017} by \cite{martin2019} suggests that LSBs start as HSBs before $z \sim 2$, but through supernova feedback and tidal interactions, the density of the galaxy is decreased, making it more diffuse and dimmer in surface brightness. Another analysis by \cite{sales2020} identifies tidal interactions and stripping as an evolutionary pathway for ultra-diffuse galaxies (UDGs), a sub-class of LSBs, to form. In dense environments, these processes are assisted by ram pressure stripping. 

The environments of LSBs are diverse; they are found to be common in clusters \citep{koda2015}, but they have been observed in all environments \citep{merritt2016, papastergis2017, roman_trujillo2017, greco_etal2018}. Compared to the HSB population, LSBs appear to be quenched, and mostly occupy the red sequence \citep{van_dokkum2015, van_dokkum2016, ferre_mateu2018, ruiz_lara2018}. This observational evidence, along with predictions from hydrodynamical simulations, suggest that there is not one characteristic evolutionary mechanism alone which is responsible for the formation of LSB galaxies.

The study of LSBs is important in a cosmological context because many current problems in the $\Lambda$ Cold Dark Matter ($\Lambda$CDM) model, such as overmerging \citep{klypin1999}, cold collapse \citep{moore1999}, and subhalo darkness \citep{boylan_kolchin2011} may be resolved by studying LSBs. In the $\Lambda$CDM model, LSBs naturally arise from dark matter halos with high angular momentum \citep{dalcanton1997}, which is supported by some observational data \citep{jimenez1998}. However, recent simulations seem to suggest that LSBs have similar angular momentum to HSB galaxies \citep{martin2019}. Nevertheless, the importance of LSBs in cosmological models, along with their general paucity in past surveys \citep{ferguson_mcgaugh1994, williams2016, kaviraj2020, kim2023}, indicates that studying properties of the LSB population can provide a more complete picture of galaxy formation.  

Because of the difficulty in finding large samples of LSBs, the majority of studies have been limited to observing those in the local Universe \citep[$z \lesssim 0.15$,][]{geller2012, greene2022}. Prior observational studies have suffered from primarily being ground-based \citep{rosenbaum_bomans2007, venhola2017, zaritsky2024}, which limits the resolution and available wavelengths, restricting the redshift and luminosity of the observed LSB populations. With the launch of JWST \citep{gardner_JWST2023}, we can now observe LSBs at longer wavelengths with higher resolution and improved sensitivity than before, enabling the calculation of more precise photometric redshifts. Central to this effort is JWST's Near Infrared Camera (NIRCam) providing a wavelength range of $0.6$ to $5.0 \; \mu m$ \citep{rieke2023a}. 

In this work, we develop a method for selecting LSBs beyond the local Universe ($0.4 < z < 0.8$) in the JADES (JWST Advanced Deep Extragalactic Survey) GOODS-S (Great Observatories Origins Deep Survey-South) field by using photometric redshifts and measuring surface brightness through the image fitting programme \texttt{pyimfit} \citep{erwin2014}. We then measure photometrically-derived properties of the sample through SED fitting, and compare these $0.4 < z < 0.8$ LSBs with other objects in the JADES GOODS-S field, particularly at lower redshift ($z < 0.4$) LSBs and HSBs with similar stellar masses and redshifts. These comparisons are made to investigate and test previous theories on LSB formation and evolution. 

This paper is outlined as follows. In Section \ref{sec:data}, we describe the data and photometric catalogue used to conduct this survey. In Section \ref{sec:selection}, we describe the process used to select objects in our $0.4 < z < 0.8$ LSB sample. In Section \ref{sec:LSB_properties}, we measure the spatial distribution, number density, multi-band surface brightness profiles, and colours of the LSBs in this sample. In Section \ref{sec:LSB_comparisons}, we derive the star formation histories, spectral energy distributions, mass-weighted ages, and dust attenuations for LSBs in this sample, comparing them with lower redshift LSBs and brighter objects at similar redshift. Section \ref{sec:conclusion} presents the findings from these properties and comparisons.  

All magnitudes used in this paper use the AB system \citep{oke_gunn1983}. Uncertainties are quoted as 68\% confidence intervals, unless otherwise stated. Throughout this work, we report vacuum wavelengths and adopt a standard flat $\Lambda$CDM cosmology from Planck18 \citep{Planck_collab2020} ($H_0 = 67.4$ km s$^{-1}$ Mpc$^{-1}$, $\Omega_{\rm m} = 0.315$).

\section{Data and Photometric Catalogue} \label{sec:data}

This work uses deep optical imaging from the Advanced Camera for Surveys (ACS) on the Hubble Space Telescope (HST) alongside deep infrared imaging from JWST/NIRCam in the GOODS-S field \citep{GOODS-S}. All data used are photometric with five optical-infrared bands from ACS (F435W, F606W, F775W, F814W, and F850LP), and 15 infrared bands from NIRCam (F070W, F090W, F115W, F150W, F182M, F200W, F210M, F277W, F335M, F356W, F410M, F430M, F444W, F460M, and F480M). These 20 filters collectively cover wavelengths from 0.4-5.0 microns. 

The HST/ACS mosaics used in this work were produced as part of the Hubble Legacy Fields (HLF) project v2.0 \citep{illingsworth2016, whitaker2019}. The JWST/NIRCam data were primarily obtained as part of the JADES \citep{eisenstein2023} programme IDs (PIDs) 1180, 1210, 1286, 1287, 3215, and 6541. In addition, photometric data from the following additional JWST programmes were used: Windhorst IDS GTO programme \citep[PID 1176;][]{windhorst2017}, Mid-Infrared Instrument (MIRI) Deep Imaging Survey \citep[MIDIS; PID 1283;][]{MIRI_HUDF2024}, First Reionization Epoch Spectroscopic COmplete Survey \citep[FRESCO; PID 1895;][]{oesch2023}, JWST Extragalactic Medium-band Survey \citep[JEMS; PID 1963;][]{jems2023}, Next Generation Deep Extragalactic Exploratory Public (NGDEEP) Survey \citep[PID 2079;][]{ngdeep2024},  \cite[PID 2198;][]{barrufet2021}, PANORAMIC survey \citep[PID 2514;][]{panoramic2025}, \cite[ALESS-JWST; PID 2516;][]{jwst_prop2516}, and BEACON \cite[PID 3990;][]{jwst_prop3990}. 

For a detailed description of the JADES JWST/NIRCam imaging data reduction and mosaicing, consult \cite{tacchella2023}. The JWST/NIRCam source detection is detailed in \cite{robertson2023}. The main steps are stacking six image mosaics (F200W, F277W, F335M, F356W, F410M, and F444W) using the corresponding error images and inverse-variance weighting to produce a single detection image. Within this detection image, a source catalogue is constructed by selecting contiguous regions of greater than 5 pixels (0.15 arcsec) with a signal-to-noise ratio (SNR) of > 3 and applying the \texttt{SExtractor} \citep{SExtractor_1996} deblending algorithm \citep{photutils2022}. Photometry convolved to the PSF of F444W is then performed at the source centroids in all of the aforementioned photometric bands from HST/ACS and JWST/NIRCam (with the exception of the F430M, F444W, F460M, and F480M bands) with circular ($0.1^{\prime\prime}-0.5^{\prime\prime}$ radius) apertures and elliptical Kron apertures (Kron parameter 1.2, i.e. ``small Kron'' and Kron parameter 2.5, i.e. ``large Kron''). Uncertainties are estimated by placing random apertures across regions of the image mosaics to compute a flux variance. To account for differing depths across the fields, these random apertures are collected into percentiles based on their location's exposure time, which are then used to determine the sky-noise contribution to the flux uncertainty for each object.

\section{Sample Selection}
\label{sec:selection}

% Putting Figure 1 here so it doesn't push all the other figures down a page
\begin{figure*}
    \centering
    \includegraphics[width=0.75\linewidth]{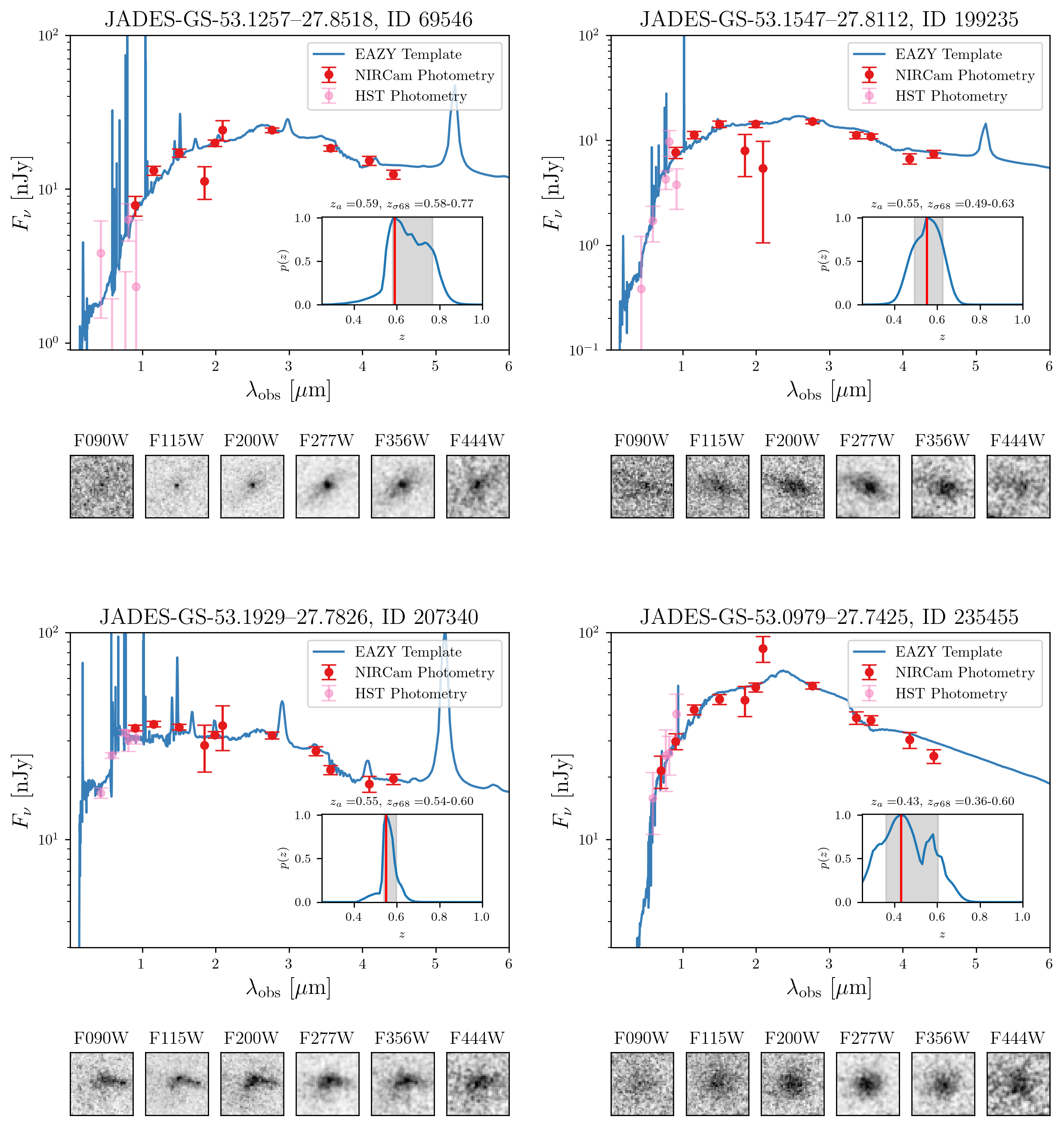}
    \caption{Example photometric redshift SED fits from \texttt{eazy-py} shown for four LSB candidates. Small Kron photometry from NIRCam filters from the catalogue described in Section \ref{sec:data} is shown in red with measured errors, while those from HST filters are shown in pink, both in nJy at the pivot wavelength (in microns) of the corresponding filter. The best-fit SED template from \texttt{eazy-py} is shown as a blue line. Each SED has an inset plot showing the probability $p(z)$ surface for the fit, where our adopted redshift $z_a$ (corresponding to the maximum $p(z)$), is shown as a red vertical line and the area between the 1$\sigma$ uncertainties of the distribution are shaded in gray. The redshift values corresponding to $z_a$ and the 1$\sigma$ spread are printed at the top of each inset plot. A mosaic cutout for each object is shown in six wide NIRCam filters below each SED.}
    \label{fig:EAZY_examples}
\end{figure*}

The primary goal of this study is to find LSB objects at higher redshifts than LSBs that have been studied before, while still resolving faint and extended objects. To achieve this, we need to consider photometric signal-to-noise, surface brightness, and photometric redshift, since surface brightness scales with redshift as $(1 + z)^{-4}$ due to cosmological dimming. 

From the full photometric catalogue described in Section \ref{sec:data}, we start by requiring that objects have a SNR of 10 or greater in the F200W filter with $0.1^{\prime\prime}$ circular aperture photometry, which yields 96,327 objects. We chose F200W (corresponding to $\sim$1.3 $\mu$m rest-frame for $z = 0.5$ galaxies and $\sim$1.0 $\mu$m rest-frame for $z = 1.0$ galaxies) as our representative filter since it has the best combination of depth and spatial resolution in JADES, with a 5-$\sigma$ point-source sensitivity of 4.4 nJy \citep{rieke2023} and spectral resolution $R= \lambda/\Delta\lambda \sim 4.3$, where $\lambda$ is the pivot wavelength and $\Delta\lambda$ is the bandwidth. This choice ensures proper coverage across NIRCam filters among objects in our sample. This step also helps to filter out any noise or artefacts that may have been falsely identified as objects in the catalogue. 

For the objects that satisfy our SNR cuts, we estimate their 2-dimensional surface brightness profiles. For this, we turn to the image fitting programme \texttt{pyimfit} \citep{erwin2014}.  We run a single component S\'{e}rsic profile fit on each of the objects with SNR(F200W) > 10 in the F200W band. Using these profiles, we can identify a sample of objects with an averaged surface brightness fainter than a certain threshold. 

The surface brightness fitting procedure goes as follows: for each object, we take a cutout of our F200W filter mosaic, centred around the object's position in the catalogue from Section \ref{sec:data} (this includes both the data and an associated one standard deviation error image). This cutout is square, with a side length of 12 times the object's catalogued semi-major axis $a$. A segmentation map is created with the cutout using a Python \texttt{SExtractor}-based package \texttt{sep} \citep{barbary2016}, to determine which pixels belong to the central source. If a pixel is determined to be a part of another object within the cutout, it is masked out from the programme's fitting routine. This is done as follows: starting with an initial threshold of 1.5$\sigma$, if any pixels are detected outside a square around the central pixel of sides 5$a$ that belong to the central object's segmentation, then the segmentation map is recreated with 0.5$\sigma$ added to the threshold until this condition is met. Visual inspection shows that this is an effective method to account for nearby objects and only fitting pixels belonging to the central object in the cutout.

The single-component S\'{e}rsic profile \citep{sersic1963} in terms of intensity $I$ as a function of disk radius $R$ goes as

\begin{equation}
    \label{eq:S\'{e}rsic_intensity_profile}
    I(R) = I_{\rm eff}\exp \left\{-b_n \left[\left(\frac{R}{R_{\rm eff}} \right )^{1/n} - 1\right ] \right\},
\end{equation}

where $R_{\rm eff}$ is the effective (or half-light) radius, $I_{\rm eff}$ is the intensity at the half-light radius, and $n$ is the S\'{e}rsic index. This index $n$ is a parameter that determines the shape of the profile, with $n=0.5$ producing a Gaussian profile, $n=1$ producing an exponential profile, and $n=4$ producing a de Vaucouleurs profile. A lower $n$ generally corresponds to a shallower profile, so for extended objects like LSBs, we expect this index to be particularly low. $b_n$ is a geometric parameter that is dependent on $n$, and is determined by the condition

\begin{equation}
    \label{eq:b_n}
    \Gamma (2n) = 2\gamma (2n, b_n),
\end{equation}
where $\Gamma$ and $\gamma$ represent the complete and incomplete gamma functions, respectively. 

The surface brightness $\mu$ goes as 

\begin{equation}
    \label{eq:S\'{e}rsic_brightness_profile}
    \mu(R) = \mu_{\rm eff} + \frac{2.5b_n}{\ln(10)}\left [ \left ( \frac{R}{R_{\rm eff}} \right )^{1/n} - 1 \right ],
\end{equation}
where $\mu_{\rm eff}$ is $\mu(R_{\rm eff})$.

This profile model has $R_{\rm eff}$, $I_{\rm eff}$, and $n$ as free parameters, all of which \texttt{pyimfit} estimates, along with the object's ellipticity and position angle (to account for its geometry and orientation in the image, respectively). The object's position is held fixed within plus or minus 5 pixels in both $x$ and $y$ coordinates of that from the catalogue in Section \ref{sec:data}. 

From the total apparent magnitude $m_{\rm{tot}}$ given by the best fit model, we can find the mean effective surface brightness (sometimes referred to as the mean surface brightness) $\bar{\mu}_{\rm{eff}}$, the average brightness within the circularised effective radius, by

\begin{equation}
    \label{eq:mean_eff_sb}
    \bar{\mu}_{\rm{eff}} = m_{\rm{tot}} + 2.5\log[2\pi (1 - \epsilon) R_{\rm{eff, arcsec}}^2] - 10\log(1 + z),
\end{equation}
where the total flux measured is scaled by $(1 + z)^4$ (third term in Eqn \eqref{eq:mean_eff_sb}) to account for the aforementioned cosmological dimming. The $(1 - \epsilon)$ term (where $\epsilon$ is the fitted ellipticity of the object) is included because the programme fits the S\'{e}rsic profile along the semi-major axis and $R_{\rm eff}$ is a circularised radius. For the redshift scaling, we derive photometric redshifts for the sample using \texttt{eazy-py}, a Python wrapper of the programme \texttt{EAZY} \citep{brammer2008}. We use an internal catalogue obtained in the same manner as in \cite{hainline2024}, using the ``small Kron'' photometry for the SED fitting. Examples of these \texttt{eazy-py} SED fits are shown in Figure \ref{fig:EAZY_examples} for four LSB candidates. The best-fit SED template, small Kron photometry, probability $p(z)$ surfaces, and mosaic cutouts in six wide NIRCam filters are shown for each candidate. For each object in this study, we adopt $z_a$, the redshift at which the $p(z)$ surface is at a maximum, as our canonical redshift.

\begin{figure*}
    \centering
    \includegraphics[width=0.75\linewidth]{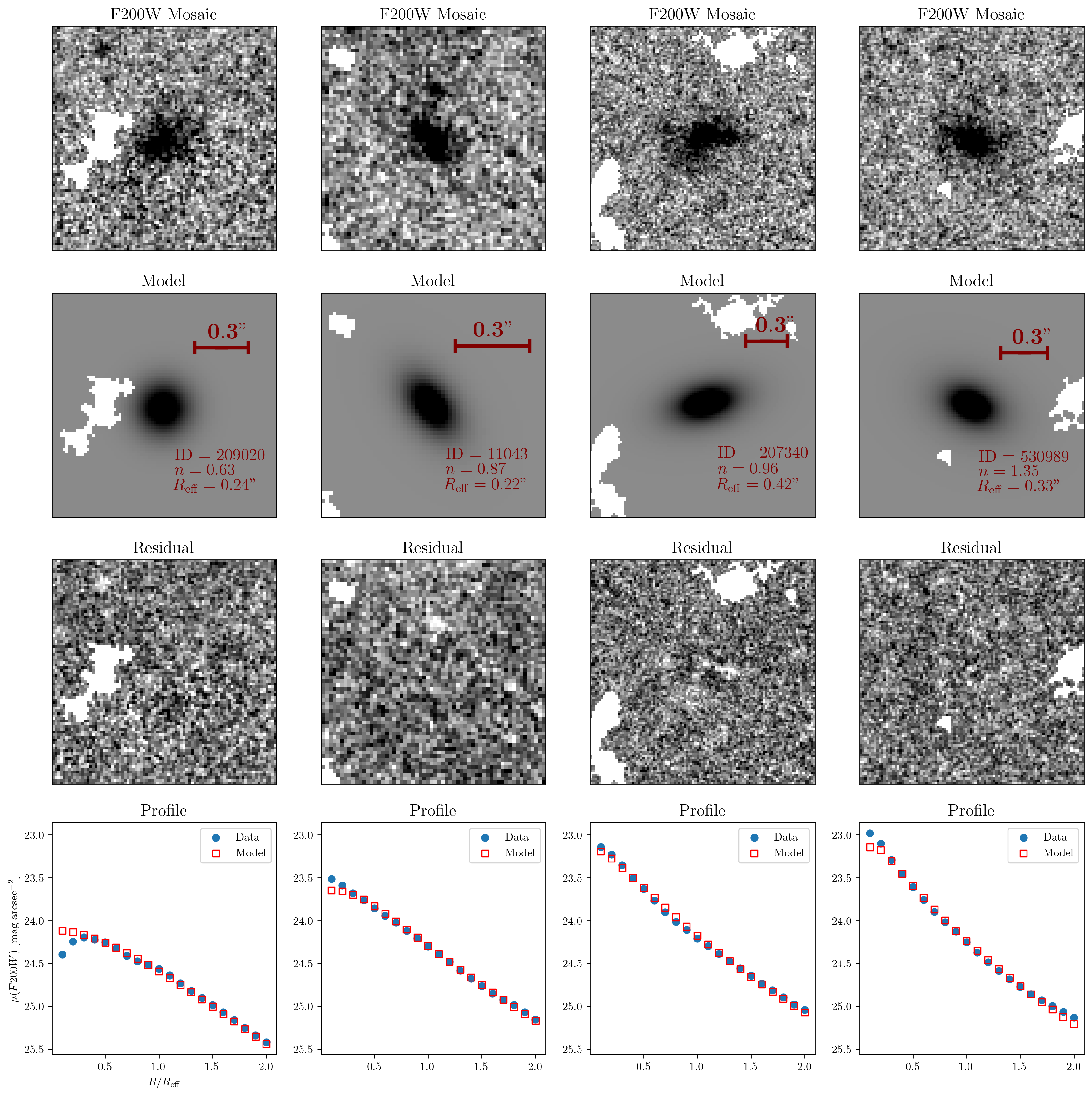}
    \caption{\texttt{pyimfit} 2D surface brightness profiles for four LSB candidates chosen randomly from bins of the S\'{e}rsic index $n$. From top to bottom, the rows show the original cutout from the F200W mosaic, the best-fit S\'{e}rsic model, the residual (data minus model), and the profile plotted against the data, by drawing many ellipses at different radii, to the object's fitted $R_{\rm eff}$. The white pixels were determined to belong to another object within the cutout, and were masked from the fits. Each object's JADES ID, $n$, and $R_{\rm eff}$ are printed in each model panel, as well as an angular scale corresponding to $0.3^{\prime\prime}$ for each object. Cutouts are 12 times the object's catalogue semi-major axis.}
    \label{fig:pyimfit_example}
\end{figure*}

From this mean effective surface brightness, we can also find the central surface brightness through

\begin{equation}
    \label{eq:central_sb}
    \mu_0 = \bar{\mu}_{\rm{eff}} + 2.5\log\left [ \frac{n{e}^{b_n}}{(b_n)^{2n}} \Gamma(2n) \right ] - \frac{2.5 b_n}{\ln(10)}
\end{equation}

For details on the derivations of these formulae, consult \cite{graham2005}. 

While a single S\'{e}rsic may not be sufficient to fit brighter, spatially resolved objects in the catalogue, it works sufficiently well for objects that are smaller, dimmer, and more likely to be LSBs. Some examples of these \texttt{pyimfit} fits on LSBs at varying values of $n$ are provided in Figure \ref{fig:pyimfit_example}. The selected objects vary in ellipticity and position angle. Each column represents an object, with the rows showing (from top to bottom) the mosaic image in F200W, the fitted model from \texttt{pyimfit}, the residual (data minus model), and the fitted surface brightness profile. The groups of white pixels in the first three rows were determined to belong to another object through the \texttt{sep} segmentation procedure and were masked from the fitting routine. The profiles on the bottom row are constructed by drawing concentric ellipses at different radii and measuring the brightness within each ellipse, both for the original image (first row) and the \texttt{pyimfit} model (second row). 

In Figure \ref{fig:pyimfit_example}, the object's value of $n$ increases from left to right (each column represents an object). ID 207340 has an irregular morphology, and the residual shows that the single-component S\'{e}rsic fit leaves noticeable residuals, though the 1D surface brightness profile is well fit by the model. Figure \ref{fig:pyimfit_example} demonstrates that the fits we perform encapsulate the properties of the S\'{e}rsic profile for the objects we are interested in, which validates the use of Equations \eqref{eq:mean_eff_sb} and \eqref{eq:central_sb} as measures of surface brightness.

Motivated by \cite{greco_etal2018}, we choose mean effective surface brightness $\bar{\mu}_{\rm eff}$ (Eqn \eqref{eq:mean_eff_sb}) as our defining parameter instead of central surface brightness (Eqn \eqref{eq:central_sb}) to allow consideration of HSB bulges in extended LSB disks. While the half-light radius $R_{\rm eff}$ can be larger than the object's physical size, some objects have a fitted $R_{\rm eff}$ that is many times larger than the catalogue semi-major axis for the source, which leads to an underestimation in the mean effective surface brightness. Therefore, we first require that objects have a fitted effective radius less than 1.5 times their catalogue semi-major axis ($R_{\rm eff} < 1.5a$) to avoid this overestimation. Since we want to focus on somewhat extended LSBs, we then require that our objects need an effective radius of at least 0.18 arcsec. 

\begin{figure}
    \centering
    \includegraphics[width=1\linewidth]{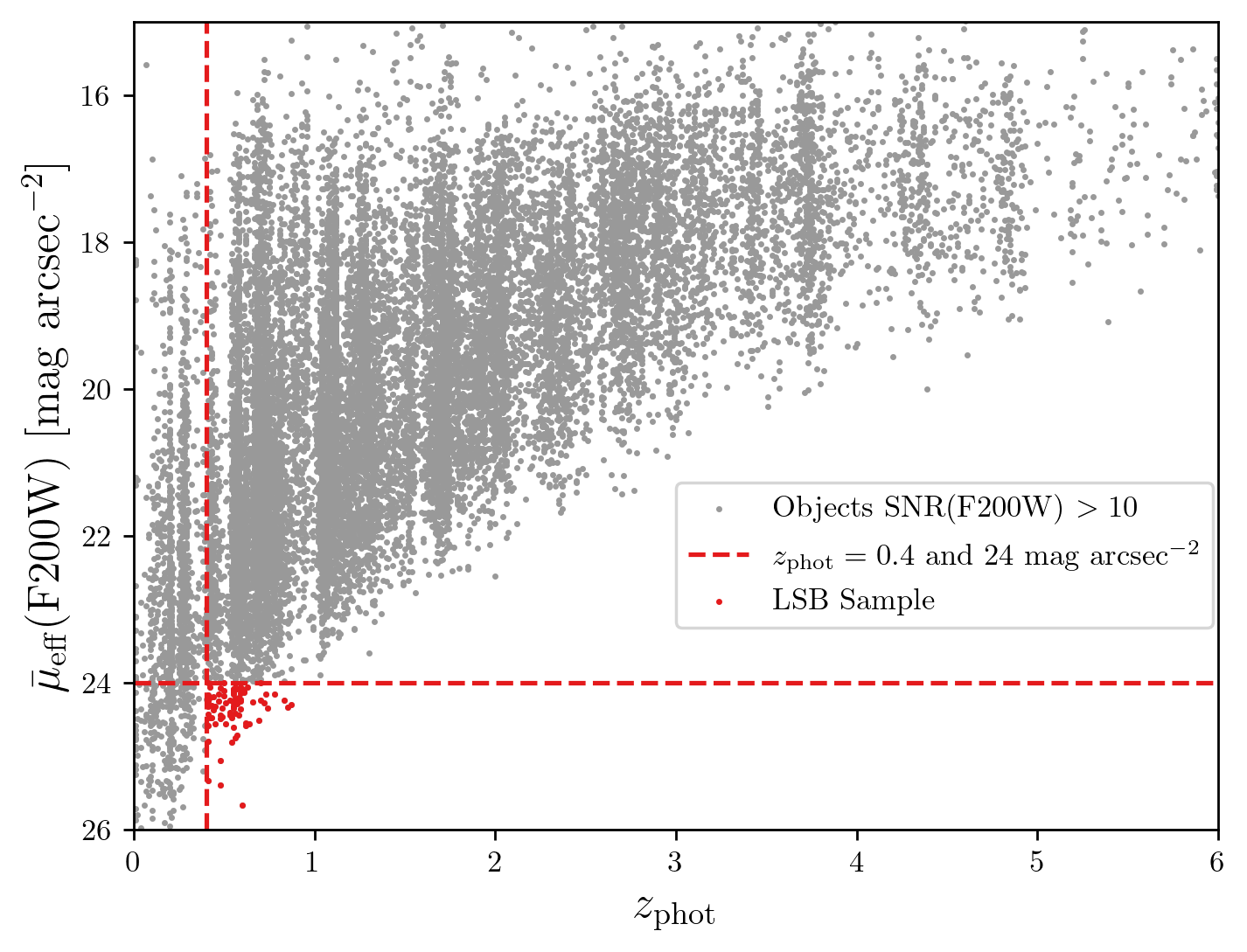}
    \caption{The mean effective surface brightness given by Eqn \eqref{eq:mean_eff_sb} from \texttt{pyimfit} in the F200W filter (corrected for cosmological dimming with photometric redshifts from \texttt{eazy-py}) against the photometric redshift from \texttt{eazy-py} for each object with SNR > 10 in F200W, fitted effective radius less than 1.5 times their semi-major axis, and fitted effective radius greater than 0.18 arcsec. A vertical dashed line is drawn at redshift $z_{\rm phot}=0.4$ and a horizontal dashed line is drawn at $\bar{\mu}_{\rm eff} = 24$ mag arcsec$^{-2}$, both in red. The LSB sample, objects $z_{\rm phot} > 0.4$ and $\bar{\mu}_{\rm eff} > 24$ mag arcsec$^{-2}$, is highlighted in red.}
    \label{fig:brightness-z}
\end{figure}

With a measure of surface brightness and redshift for each of these objects, we can explore the relationship between the two and use it to identify a sample of LSBs. In Figure \ref{fig:brightness-z}, we plot the mean effective surface brightness in F200W $\bar{\mu}_{\rm eff}(\rm F200W)$ from \texttt{pyimfit} against the photometric redshift from \texttt{eazy-py} for every object after making the SNR and $R_{\rm eff}$ cuts. 

Even with the $(1 + z)^4$ cosmological dimming taken into account in Eqn \eqref{eq:mean_eff_sb}, Figure \ref{fig:brightness-z} shows that the ability to detect objects fainter than 24 mag arcsec$^{-2}$ gets increasingly difficult with redshift. This is due to the detection limits of JWST, as well as the presence of background light due to zodiacal light and emission from the telescope which decrease the contrast that allows for detecting these faint sources. A more detailed discussion on the exposure time required to detect these objects at higher redshifts is provided in Section \ref{sec:conclusion}. Nevertheless, if we choose our brightness cut to be 24 mag arcsec$^{-2}$, we still recover objects out to $z \sim 0.8$, which is significantly farther than previous surveys \citep{geller2012, greene2022}. To select a sizeable sample where we see some evolution in these LSBs, we choose to make a redshift cut at $z_{\rm phot} > 0.4$. These cuts in F200W surface brightness and photometric redshift are shown as red lines in Figure \ref{fig:brightness-z} with the LSB sample highlighted in red. 

\begin{figure*}
    \centering
    \includegraphics[width=0.75\linewidth]{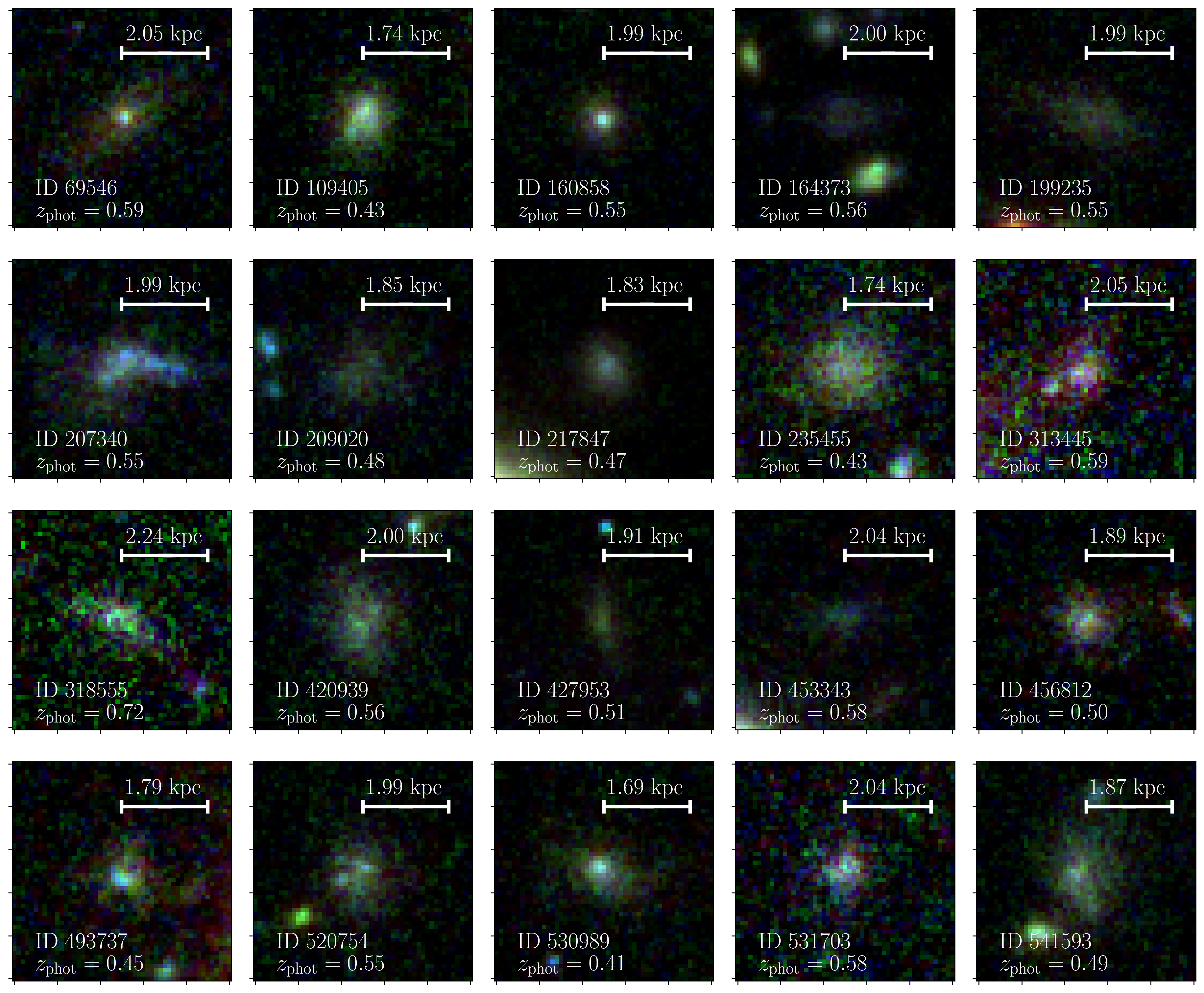}
    \caption{False-colour RGB (F277W/F200W/F115W) images for the 20 largest LSBs in the sample, by effective radii. Each cutout is 50 $\times$ 50 pixels (1.5 $\times$ 1.5 arcsec), and each object is normalised by its individual cutout's brightest pixel. Each object's JADES catalogue ID and photometric redshift are shown in the lower left corners, while the physical scale corresponding to 0.3 arcsec is provided in the top right corners.}
    \label{fig:LSB_RGBs}
\end{figure*}

Finally, we visually inspect each of these objects to identify false positives and ensure our sample is comprised of actual faint and extended galaxies. These false positives can include tidal tails, shredded galaxy subcomponents, and bits of diffraction spikes from bright stars. False-colour RGB images from the F115W, F200W, and F277W filters of many of the largest LSBs (by fitted $R_{\rm eff}$) in this sample are shown in Figure \ref{fig:LSB_RGBs} at fixed 50 $\times$ 50 pixel (1.5 $\times$ 1.5 arcsec) panels. After these visual cuts, we are left with 57 galaxies in our final LSB sample, with photometric redshifts in the range of $0.4 < z < 0.8$. Our analysis suggests JADES data cannot detect LSBs ($\bar{\mu}_{\rm eff}(\rm F200W) > 24 \; mag \; arcsec^{-2}$) past $z \sim 0.8$ in the way we have defined them. Results from \texttt{pyimfit} for these 57 LSBs are reported in Table \ref{tab:pyimfit}.

Visually, the LSBs in Figure \ref{fig:LSB_RGBs} are faint and extended. Some of these objects, like ID 530989, appear to have a bright centre surrounded by a dim disk, similar to the well-known LSBs such as Malin 1 \citep{bothun_malin1}.

\section{Properties of the LSB Sample}
\label{sec:LSB_properties}

\subsection{Photometric Redshift}
\label{subsec:zphot}
\begin{figure}
    \centering
    \includegraphics[width=1\linewidth]{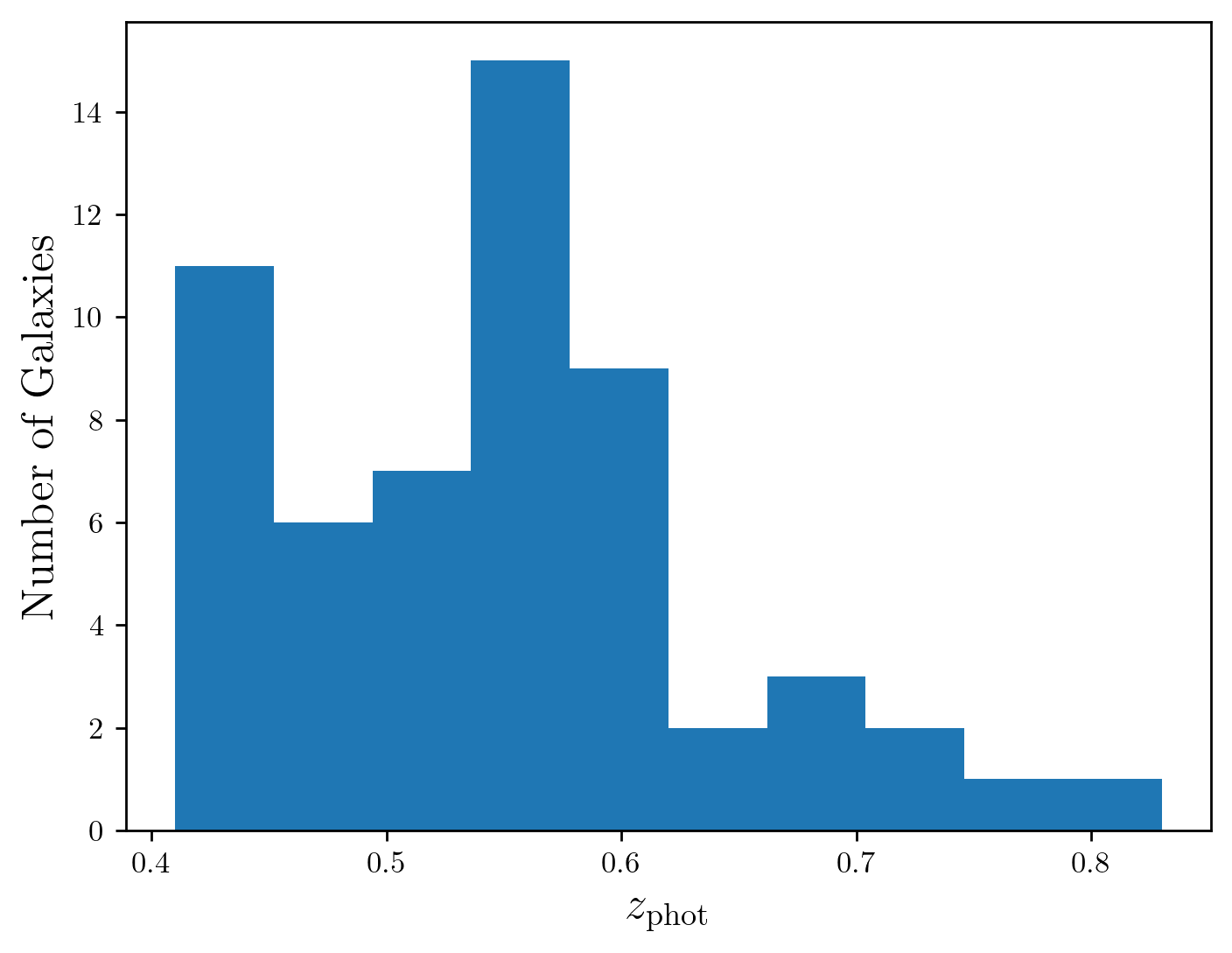}
    \caption{Distribution of photometric redshifts obtained through \texttt{eazy-py} (as described in Section \ref{sec:selection}) for our final sample of LSBs.}
    \label{fig:z_dist}
\end{figure}

We will now briefly discuss the distribution of photometric redshifts from \texttt{eazy-py} for our final sample of LSBs. As discussed in Section \ref{sec:selection} and shown in Figure \ref{fig:EAZY_examples}, these photometric redshifts were determined through an internal catalogue using the ``small Kron'' aperture. The distribution of these redshifts, ranging from $0.4 < z_{\rm phot} < 0.8$, is shown in Figure \ref{fig:z_dist}. 

\subsection{Spatial Distribution}
\label{subsec:spatial_dist}
\begin{figure*}
    \centering
    \includegraphics[width=0.75\linewidth]{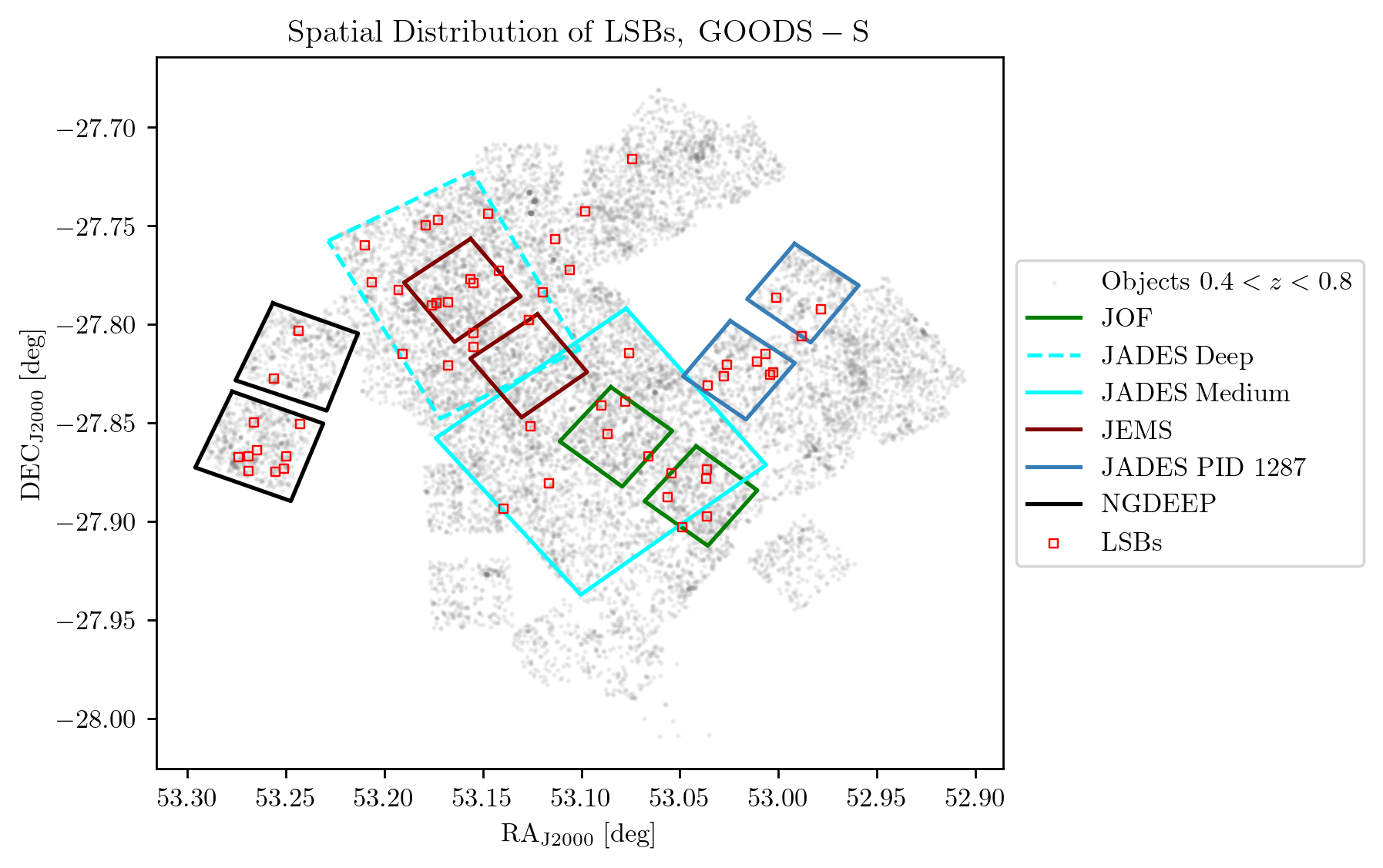}
    \caption{Positions of our LSB sample are plotted as red squares, while other objects above the SNR cut (see Section \ref{sec:selection}) in the same range of photometric redshifts are plotted in a faint grey, to better see the density of surrounding objects. It should be noted that some of the dark grey dots can correspond with faulty segmentation of debris surrounding a large galaxy. The JADES medium and deep fields, as well as the JOF, JEMS, PID 1287 and NGDEEP (see Section \ref{sec:data}) coverage are highlighted in different colours.}
    \label{fig:LSB_spatial_dist}
\end{figure*}

In Figure \ref{fig:LSB_spatial_dist}, we show the spatial distribution of the LSB sample across GOODS-S as red squares, while every other object with SNR$>10$ in F200W from the same redshift range ($0.4 < z_{\rm phot} < 0.8$) is represented by a faded grey dot (to show dense regions of the field). The boxes in the figure identify the JADES regions of greater depth, as well as outlines of other areas of interest in GOODS-S.

The LSBs are not uniformly distributed; they generally are more abundant in the deeper parts of the field, which is a selection effect. Most of these LSBs appear in regions of GOODS-S with deep coverage: out of the 57 galaxies in our sample, 18 can be found in the JADES Deep field, 11 in the NGDEEP survey, 10 in the JADES Origins Field \citep[JOF;][]{JOF}, and 10 in JADES PID 1287, all of which contain the deepest data in this field. This implies that survey depth is a strong driver of finding LSBs at higher redshift.

While we cannot make any statistical conclusions about any physical clustering of LSBs from the size of this sample, previous studies on LSBs have found that, physically, they are not uniformly distributed \citep{koda2015, greco_etal2018, roman2021} and that denser environments may aid in LSB formation due to ram pressure stripping \citep{martin2019}. However, it has been established that LSBs occur in all environments, so it is not surprising to see some of our LSBs in less dense environments. Given the varying depth across the GOODS-S field, it is not possible to reach a robust conclusion on how the distribution of JADES LSB galaxies at $z > 0.4$ correlates with environment.

\subsection{Number Density}
\label{subsec:num_dens}
The number density of LSBs has been previously studied and reported by numerous papers. \cite{dalcanton_num_dens1997} uses a wide scale CCD survey to measure the number density of LSBs with $23 < \mu_0(V) < 25$ mag arcsec$^{-2}$, reporting $\sim 4$ galaxies per square degree. Correcting for the completeness of LSBs of smaller size, \cite{greene2022} follows up on this analysis by exploring the densities at fixed mass using the HSC-SSP survey. They find that the number densities of low redshift LSBs are consistent with known numbers around $M_* \sim 10^7 M_{\odot}$, but grow inconsistent at higher masses. There is also some evidence that number density increases with decreasing mass \citep{dalcanton1995, sedgwick2019, kim2023}.

For our LSB sample, we can do a rough calculation of the number density of LSBs in GOODS-S. The total survey area of the JADES GOODS-S field considered here is $\sim$200 arcmin$^{2}$. Using the comoving distance between $z \sim 0.4$ and $z \sim 0.8$ to approximate the total volume searched, we estimate the comoving volume of the area searched to be $\sim 43,910$ Mpc$^{3}$. Using this comoving volume estimation, we find that for our 57 LSBs, the number density of these objects in the way we've defined them as $\mu_{\rm eff}(\rm F200W) > 24$ mag arcsec$^{-2}$ is on the order of $\sim 10^{-3}$ Mpc$^{-3}$. This is consistent with the LSB density values reported in \cite{greene2022} for the range of $M_* \sim 10^7-10^8 M_{\odot}$, which is roughly the range of stellar masses for our LSB sample (see Figure \ref{fig:Av_stellar_mass}). 

However, as mentioned in Section \ref{subsec:spatial_dist}, the spatial distribution of our LSB sample is biased towards the areas of the survey that are of greatest depth. We can consider the LSBs in the survey areas with the greatest depth (NGDEEP, JOF) and calculate the number density by using the angular area of the polygons drawn in Figure \ref{fig:LSB_spatial_dist}. This can ensure an approximately constant depth. There are 11 of our LSBs in NGDEEP and 10 of our LSBs in the JOF (see Figure \ref{fig:LSB_spatial_dist}). Using these numbers, we find that both number densities are also on the order of $\sim10^{-3}$ Mpc$^{-3}$.

\subsection{Surface Brightness Profiles (Multi-Band)}
\label{subsec:pysersic_profiles}
As described in Section \ref{sec:selection}, the LSBs in our sample underwent an initial round of surface brightness fitting with \texttt{pyimfit} in the F200W band. This first round of fitting is run on all the objects in our catalogue with SNR $> 10$ in F200W to measure relative surface brightness, and PSF broadening is not accounted for. The initial fits are for picking out the faintest, most extended objects past $z_{\rm phot} > 0.4$. We further perform a more deliberate, PSF-convolved, accurate S\'{e}rsic profile fit in multiple filters, for which we employ \texttt{pysersic} \citep{pasha_miller2023}. \texttt{pysersic} is more recently developed in contrast to \texttt{pyimfit}, utilizes Bayesian inference for speed, and is specialised for single-component S\'{e}rsic fits. The purpose of this second round of fitting is to gather accurate S\'{e}rsic profile parameters in different filters for each of our LSBs, as well as measuring the surface brightness in multiple filters that take PSF convolution into account. At a base level, \texttt{pysersic} operates much in the same way as \texttt{pyimfit} does, with similar fit results as what is shown in Figure \ref{fig:pyimfit_example} for \texttt{pyimfit}. A direct comparison figure is provided in Appendix \ref{app:pyimfit_pysersic}.

We will be using the same procedure described in Section \ref{sec:selection}: we take a 12$a$ square cutout for each object ($a$ being the catalogue semi-major axis for the object) in the data images and associated error images. \texttt{pysersic} fits for the same free parameters as \texttt{pyimfit} for the single-component S\'{e}rsic-profile ($R_{\rm eff}$, $I_{\rm eff}$, $n$, $\epsilon$, and position angle). We keep the same limits on each of the profile parameters: the central $x$ and $y$ values are allowed to vary within 5 pixels of the object's catalogue position, but two additional constraints are added: the effective radius $R_{\rm eff}$ is limited to 1.5$a$ (a limit placed on the \texttt{pyimfit} results in Section \ref{sec:selection}), and we limit the $n$ value within the range of the previous \texttt{pyimfit} results ($0.65 < n < 3.00$). For a more in-depth description of the fitting procedure and the mathematical calculations made with the S\'{e}rsic profile parameters, see Section \ref{sec:selection}. We repeat this procedure in four different representative filters (F150W, F200W, F277W, and F356W) to gain some insight into how these S\'{e}rsic parameters change with wavelength. 

\begin{figure*}
    \centering
    \includegraphics[width=1\linewidth]{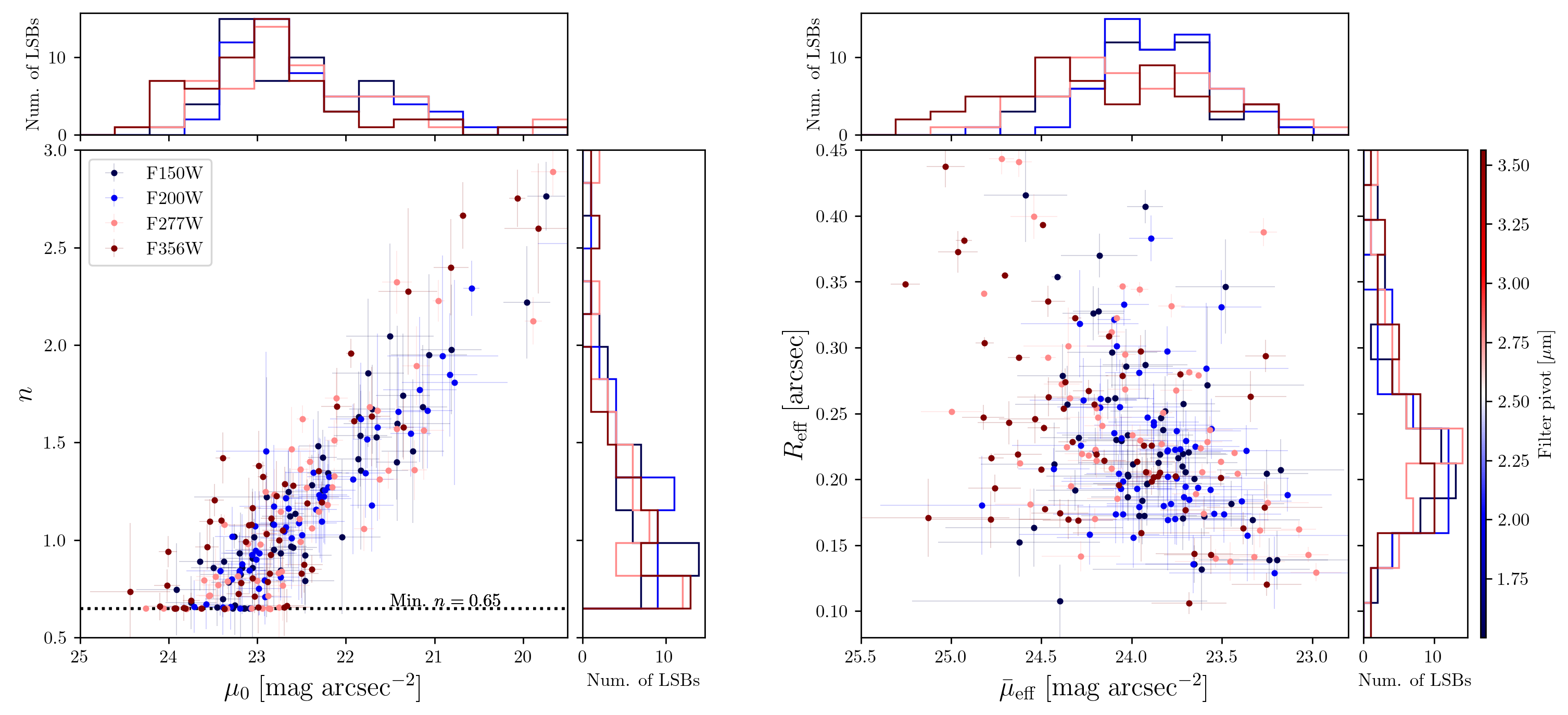}
    \caption{Distributions of the best fit S\'{e}rsic surface brightness profiles in the F150W, F200W, F277W, and F356W bands. Data points are coloured by filter as noted in the legend. S\'{e}rsic parameters are shown in scatter plots with histogrammes on each axis. Each point represents one LSB in one band. \textbf{Left:} S\'{e}rsic index $n$ against central surface brightness $\mu_0$. The lower limit of $n=0.65$ is shown as a black dotted line, while the upper limit of $n=3.00$ (imposed by pre-selection) is the upper limit of the y-axis. \textbf{Right:} Effective radius $R_{\rm eff}$ against mean effective surface brightness $\bar{\mu}_{\rm eff}$.}
    \label{fig:multi_band_params}
\end{figure*}

The best fit S\'{e}rsic parameters, mean effective surface brightness, and central surface brightness for all of the different bands are shown in Figure \ref{fig:multi_band_params} and are reported in Table \ref{tab:pysersic}. The left panel of Figure \ref{fig:multi_band_params} has the S\'{e}rsic index $n$ plotted against the central surface brightness $\mu_0$, while the right panel has the effective radius $R_{\rm eff}$ plotted against the mean effective surface brightness $\bar{\mu}_{\rm eff}$. In both panels, each data point represents an object in a given filter. The colour of the points indicates which filter was used for the parameter fitting, as noted in the legend in the left panel and the colourbar on the right. Additionally, each axis has a histogramme showing the distribution of that parameter in each filter. These distributions are coloured in the same way as the data points and each have the same number of bins. \cite{graham2013} shows a similar relation between S\'{e}rsic index $n$ against central surface brightness for a sample of elliptical galaxies regardless of surface brightness.

As expected, we see a correlation between $n$ and $\mu_0$ in the left panel of Figure \ref{fig:multi_band_params} since $n$ controls the ``steepness'' of the surface brightness profile (see Eqn \eqref{eq:central_sb}). Most LSBs in the sample have extended surface brightness profiles, so many of them stack up against the lower limit of $n=0.65$. We expect low values of $n$ for most extended LSBs \citep{koda2015}, and most of the sample has $n < 2$. However, this result may also be due to a limitation of surface brightness profile fitting as the angular size of our targets get smaller with higher redshift. As the left panel y-axis histogrammes show, the distributions for $n$ remain largely unchanged with wavelength.

The majority of the LSBs have central surface brightnesses $\mu_0$ > 22 mag arcsec$^{-2}$ across the filters, though many of them seem to have HSB ($\mu_0$ < 22 mag arcsec$^{-2}$) centres. This is a result of our choice of using $\bar{\mu}_{\rm eff}$ as our defining measure of brightness, as the distribution of $\bar{\mu}_{\rm eff}$ on the right of Figure \eqref{fig:multi_band_params} doesn't show any object that is brighter than $\sim 23$ mag arcsec$^{-2}$ (averaged over its fitted effective radius) in any band. Overestimation of $R_{\rm eff}$ can also lead to this disparity in effective and central surface brightness, as this results in averaging brightness over an area that is bigger than the object itself.

To explore possible correlations between $R_{\rm eff}$ and $\bar{\mu}_{\rm eff}$ in the different filters, we run the Spearman correlation test. This gives p-values of 0.0069, 0.0265, 0.0008, and 0.0006 for the F150W, F200W, F277W, and F356W filters, respectively. The majority of objects in most bands are between $\bar{\mu}_{\rm eff} \sim$ 23 to 25 mag arcsec$^{-2}$, which is largely consistent with our \texttt{pyimfit} single-band results. However, all of these objects were selected based on having $\bar{\mu}_{\rm eff} > 24$ mag arcsec$^{-2}$ in the F200W filter with the \texttt{pyimfit} results, but in the results for \texttt{pysersic}, there is scatter around this number, with many objects being brighter than 24 mag arcsec$^{-2}$. This is due to the effects of PSF convolution (see Appendix \ref{app:pyimfit_pysersic}), and shows that the original, unconvolved 24 mag arcsec$^{-2}$ cut in F200W translates to a roughly $\bar{\mu}_{\rm eff} \gtrsim 23$ mag arcsec$^{-2}$ cut in these four convolved bands. All LSBs in the sample are larger than $R_{\rm eff} > $ 0.18 arcsec by our selection criteria with \texttt{pyimfit}, but that requirement was purely for sample selection, so for \texttt{pysersic} we allow $R_{\rm eff}$ to freely vary up to 1.5$a$. Most objects vary in $R_{\rm eff}$ from $\sim 0.15$ to $\sim 0.45$ arcsec. At $z = 0.6$, a typical redshift for this sample, an angular distance of 0.15 arcsec corresponds to a physical distance of $\sim 1$ kpc with our adopted cosmology. 

The errorbars in Figure \eqref{fig:multi_band_params} represent the 1$\sigma$ uncertainties in the posterior estimates from \texttt{pysersic} for $n$ and $R_{\rm eff}$, which are then propagated through the calculations made for $\mu_0$ and $\bar{\mu}_{\rm eff}$. Such errors were not considered for \texttt{pyimfit} (Section \ref{sec:selection}), so this second round of fits demonstrates the difficulty of fitting the objects we are interested in for this study and measuring the surface brightness parameters accurately.

\subsection{Colours}
\label{subsec:colours}

\begin{figure}
    \centering
    \includegraphics[width=1\linewidth]{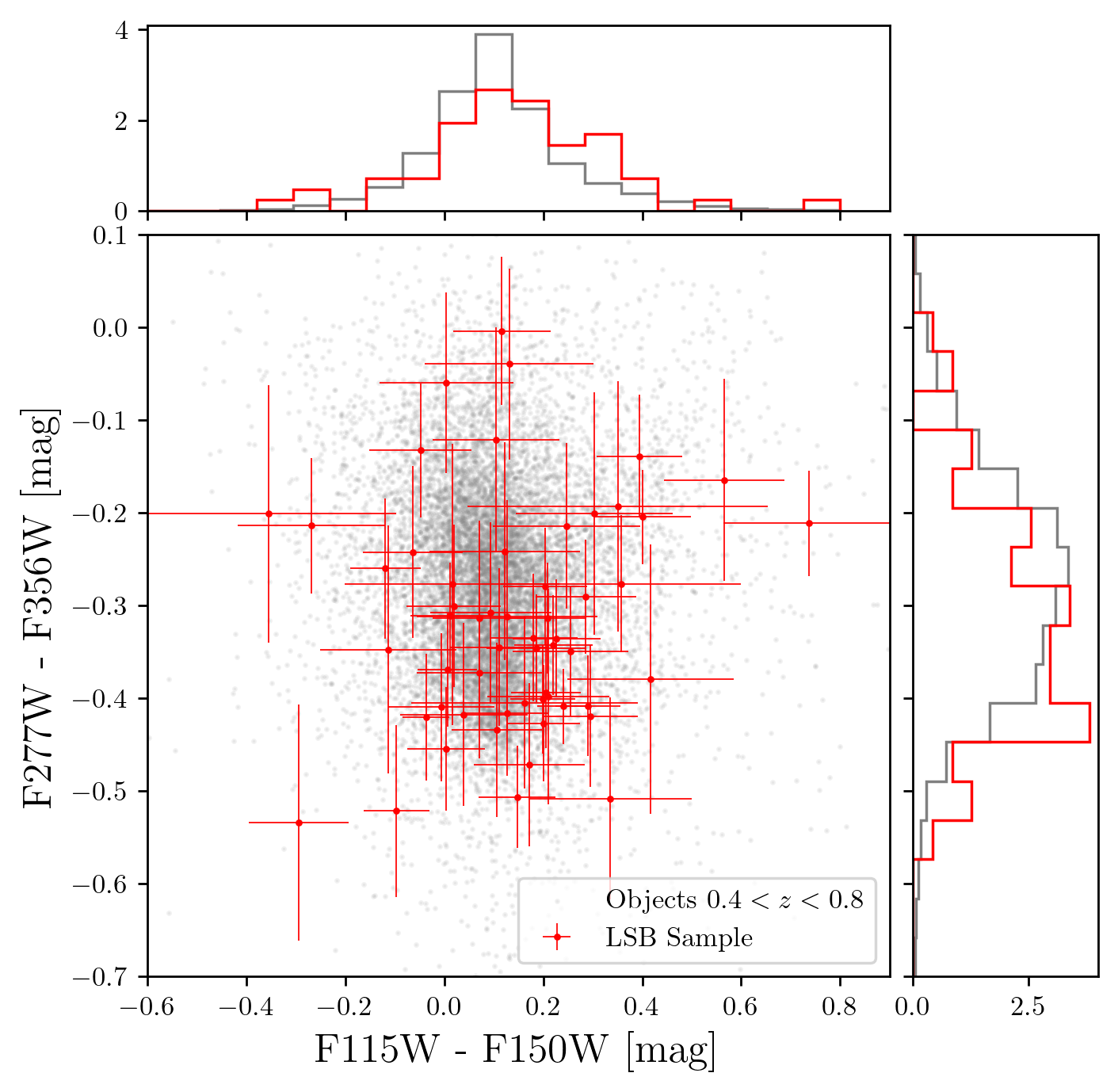}
    \caption{A colour-colour scatter plot with normalised histogrammes using the measured flux densities and errors from the catalogue described in Section \ref{sec:data} in the ``small Kron'' aperture. This figure compares the magnitude difference in the F277W and F356W filters to the difference in the F115W and F150W filters. All objects $0.4 < z_{\rm phot} < 0.8$ (reflecting the redshift range of the LSBs) are shown in light grey, while the LSB sample is shown in red. Errorbars are given for the LSB data points.}
    \label{fig:color_color}
\end{figure}

We present a colour-colour analysis of these LSBs to explore how their colours compare to those measured for objects in the rest of the field in the same redshift domain. For this, we use data directly from the photometric catalogue described in Section \ref{sec:data}, in the ``small Kron'' aperture. In Figure \ref{fig:color_color}, we take the measured differences between the F277W and F356W filters as well as the F115W and F150W filters in the ``small Kron'' aperture to construct a colour-colour plot. All objects at $0.4 < z_{\rm phot} < 0.8$ are plotted in light grey, while our LSB sample is plotted in red with error bars based on the errors given in the photometric catalogue. Histogrammes are also shown for both groups of objects on either axis, normalised as probability densities. 

With the chosen filters, the LSB sample spans the whole range of colours reflected by the overall distribution of objects at $0.4 < z_{\rm phot} < 0.8$. Previous studies have found that LSBs can be optically blue or red \citep{o'neil1997} with the redder ones being found in denser environments \citep{zaritsky2019}. Performing the two-sample Kolmogorov-Smirnov (KS) test on the LSB and overall colour distributions in both filters yields $p=0.021$ for F115W-F150W and $p=0.006$ for F277W-F356W, both of which corresponding to a 1$\sigma$ significance. This hints at a colour gradient in our LSB sample, where an LSB is more likely to be redder in F115W-F150W colour and bluer in F277W-F356W colour than other galaxies at similar redshift.

\section{Comparison of LSBs with Other Samples}
\label{sec:LSB_comparisons}

We want to see how quantities that we can photometrically derive using the programme \texttt{bagpipes} \citep{BAGPIPES_carnall2018} for our $z_{\rm phot} > 0.4$ LSBs compare with other objects in the JADES GOODS-S field. For this, we choose to make two comparisons: one with objects of similar stellar mass and redshift but higher surface brightness, and one with other LSBs at lower redshifts in the JADES GOODS-S survey ($z_{\rm phot} < 0.4$). The goal of these comparison samples is to investigate the differences in evolution between of objects with different surface brightnesses and LSBs at different redshifts. 

\subsection{Comparison Sample Selection}
\label{subsec:comp_samples}

For the HSB comparison sample of similar stellar mass and redshift, we return to the objects whose selection we discussed in Section \ref{sec:selection} that are SNR $> 10$ in F200W and at $0.4 < z_{\rm phot} < 0.8$. We take the results from \texttt{pyimfit} and use the total (redshift scaled) magnitude in F200W ($m_{\rm tot} - 10 \log (1 + z)$ from Eqn \ref{eq:mean_eff_sb}) as a proxy for stellar mass. The $z_{\rm phot} > 0.4$ LSBs have a range of 27.0-24.5 mag in this scaled total magnitude, so we select 60 random galaxies in this same range in magnitude but with a mean surface brightness of $\bar{\mu}_{\rm eff} < 23$ mag arcsec$^{-2}$.

For the low redshift LSB comparison sample, we repeat the same procedure outlined in Section \ref{sec:selection}, except instead of taking objects with $0.4 < z_{\rm phot} < 0.8$, we take all objects with $z_{\rm phot} < 0.4$. Like the HSB sample, we also require this sample to have a scaled magnitude of 27.0-24.5 mag to reflect the stellar mass distribution of the $0.4 < z_{\rm phot} < 0.8$ LSB sample and we pick 60 random galaxies that meet this criteria. We decide to make a distinction between LSBs at \$z < 0.4\$ and LSBs  at \$z > 0.4\$ to emphasize the LSBs that show evolution out to redshifts where LSBs $> 24$ mag arcsec $^{-2}$ could not be detected before.

For the photometrically derived quantities of interest, we run \texttt{bagpipes} using the ``small Kron'' photometry for each source. \texttt{bagpipes} performs SED fitting, assuming many different stellar population models, to estimate the star formation history, stellar mass, and age of a given galaxy, among other quantities. For this study, we report the best-fit SEDs, the star formation rate (SFR) averaged over the last 100 Myr $\rm SFR_{100}$, the mass-weighted age $t_{\rm MW}$, the stellar mass $M_*$,  the dust attenuation $A_V$, and the star formation history (SFR as a function of time). 

For each of these derived quantities, we assume a standard Calzetti dust model \citep{calzetti2000} with a delayed tau star-formation-history model \citep{carnall2019}, which is described by 

\begin{equation}
    \textup{SFR}(t) = Ate^{-t/\tau},
\end{equation}

where $\tau$ is a timescale that is fit for by \texttt{bagpipes}, $t$ is the time since star formation began for the object, and $A$ is a constant. The redshift is fixed to the \texttt{eazy-py} redshift described in Sections \ref{sec:selection} and \ref{subsec:zphot}, so it is not fit for by \texttt{bagpipes}. Nebular emission is modelled using the methodology of \cite{byler2017} with the \texttt{CLOUDY} photoionization code \citep{cloudy2017}. The gas phase metallicity is implicitly assumed to be the same as the stellar metallicity. 

The above assumptions made about the star formation history and the initial mass function will directly affect the derived stellar masses. However, we argue that since we are primarily doing a comparison between our three samples of galaxies using the same series of assumptions, our comparative framework is robust to choices of star formation history and the initial mass function. We also note that some of our derived stellar population properties are degenerate with one another and difficult to determine definitely with photometry alone. The most notable degeneracies are between stellar mass and stellar age in addition to between stellar metallicity and dust attenuation \citep{BAGPIPES_carnall2018, meldorf2024}.

\begin{figure*}
    \centering
    \includegraphics[width=1\linewidth]{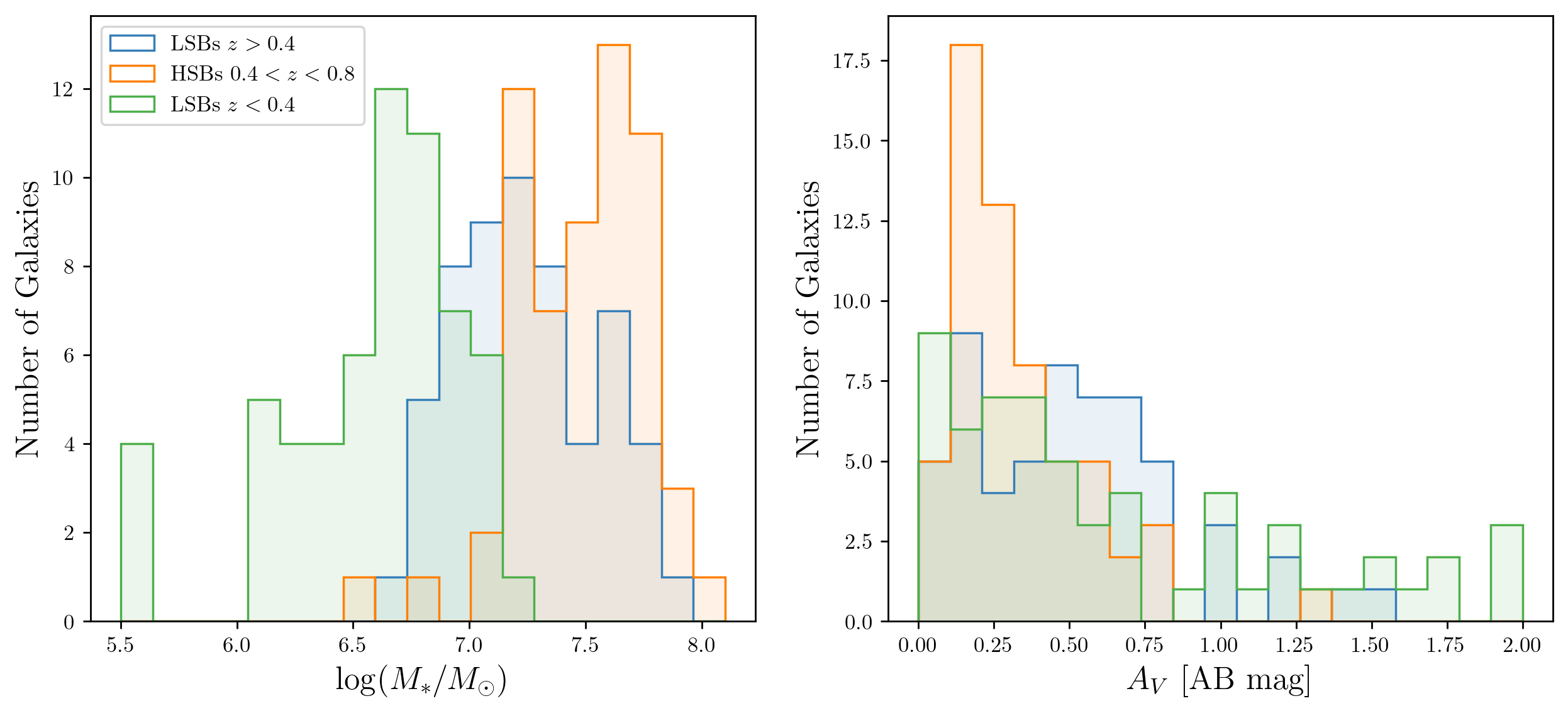}
    \caption{Histogrammes of the posterior medians of stellar mass given in $\log (M_* / M_{\odot})$ (left) and the dust attenuation $A_V$, given in AB magnitudes (right), derived from \texttt{bagpipes} for each sample. The $0.4 < z < 0.8$ LSBs are coloured in blue, the $0.4 < z < 0.8$ HSBs are coloured in orange, and the $z < 0.4$ LSBs are coloured in green.}
    \label{fig:Av_stellar_mass}
\end{figure*}

\subsection{Stellar Mass and Dust Attenuation}
\label{subsec:stellar_mass_Av}
The first photometrically-derived quantities we want to study from \texttt{bagpipes} are stellar mass $M_*$ and dust attenuation $A_V$. Stellar mass can highlight the differences in star formation between the LSBs and HSBs, while dust attenuation can tell us if light is being obscured, and if dust could be another explanation for why LSBs are low surface brightness. Histogrammes of stellar mass (left) and V-band dust attenuation $A_V$ (right) are shown in Figure \ref{fig:Av_stellar_mass}.

In general, we can see that the left panel of Figure \ref{fig:Av_stellar_mass} validates our choice of F200W magnitude as a prior for stellar mass, as all three samples lie in the dwarf ($M_* \lesssim 10^8 M_{\odot}$) regime. While there is significant overlap between the distributions of stellar mass for the three samples, Figure \ref{fig:Av_stellar_mass} shows a trend of increasing stellar mass from the $z < 0.4$ LSBs to the $z > 0.4$ LSBs to the HSBs, being partly a selection effect of the brightness criterion, with most of the $z > 0.4$ LSBs have higher stellar masses than the $z < 0.4$ LSBs. 

The right panel of Figure \ref{fig:Av_stellar_mass} shows the different distributions of V-band attenuation $A_V$ for each sample, which have significant overlap. Only the low redshift LSB sample seems to span the range of 0 to 2 magnitudes allowed by our \texttt{bagpipes} model. The HSBs seem to have noticeably lower attenuation and both LSB samples seem to have comparable attenuation. To estimate the significance level of this difference, we run the two-sample Kolmogorov-Smirnov (KS) test on these distributions. 

The KS tests involving the $A_V$ distribution for the low redshift LSBs gives $p$-values of 0.116 and 0.008 for the high redshift LSBs and HSBs, respectively. From this, we gather that the low redshift LSB distribution is different from the HSB $A_V$ distribution with $> 3\sigma$ confidence, while it is only different from the high redshift LSB distribution by $> 1\sigma$. The KS test for the $z > 0.4$ LSBs and HSBs gives a $p$-value on the order of $10^{-4}$, which corresponds to $>3\sigma$ confidence. From this, we can conclude that the distribution of dust attenuation in the low redshift LSB sample is similar to that of the higher redshift LSB sample, which is statistically different than the higher redshift HSB sample. Dust attenuation may be a factor in explaining why some of our LSB samples have lower surface brightnesses than the HSB sample, but many of the galaxies from both LSB samples have $A_V < 1.0$ mag. 92\% of the $z > 0.4$ LSBs and 73\% of the $z < 0.4$ LSBs show $A_V < 1.0$. Therefore, for the majority of our LSBs, Figure \ref{fig:Av_stellar_mass} shows that dust attenuation is not a major cause of their low surface brightness. Results and uncertainties in quantities shown in Figure \ref{fig:Av_stellar_mass} are reported in Table \ref{tab:BAGPIPES}.

\begin{figure}
    \centering
    \includegraphics[width=1\linewidth]{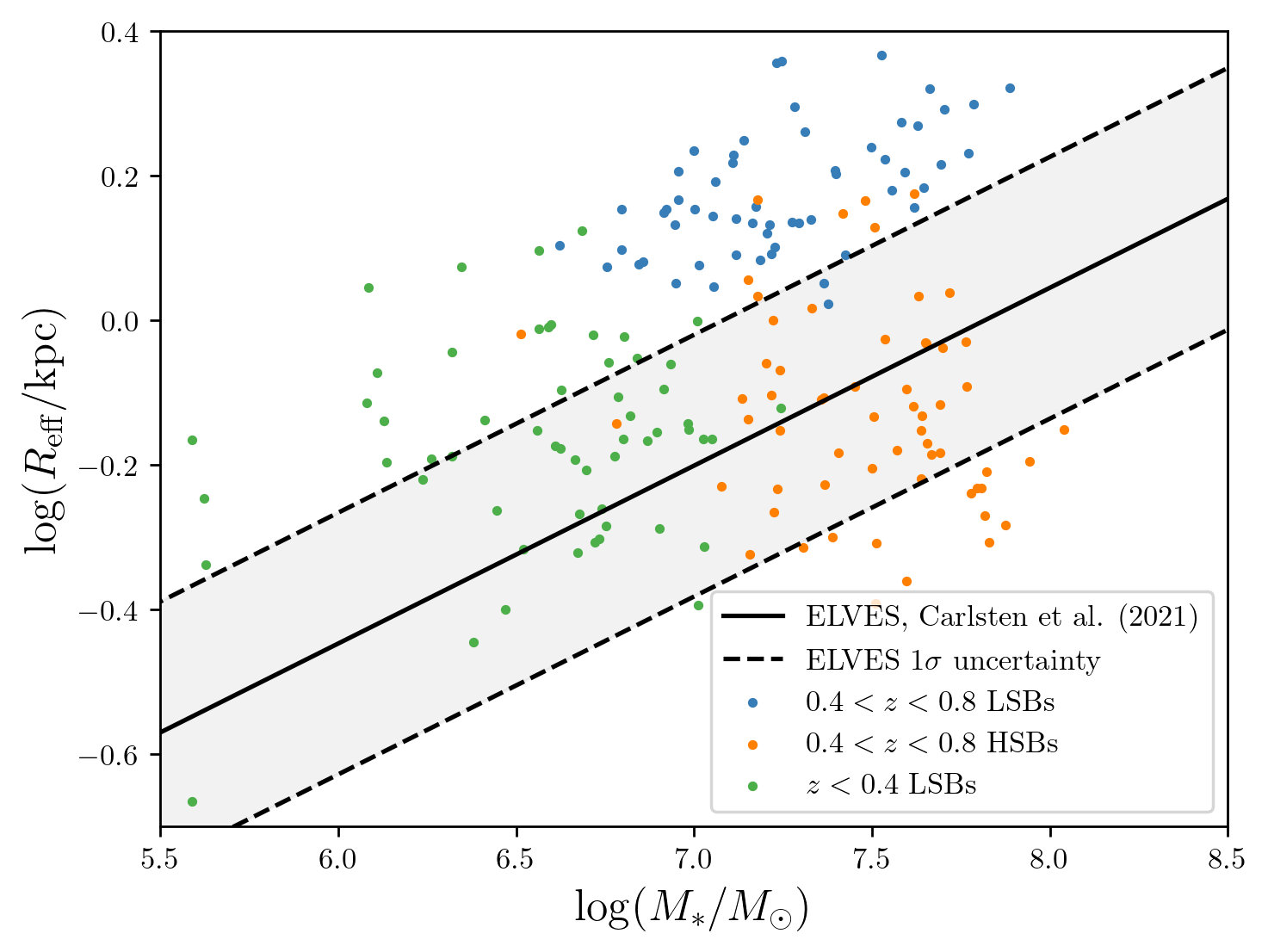}
    \caption{Size-mass relation for all three of our selected galaxy samples. The best fit effective radius from \texttt{pyimfit} is plotted as $\log(R_{\rm eff} / \rm kpc)$ against the posterior median stellar mass from \texttt{bagpipes} is plotted as $\log(M_* / M_{\odot})$. The best fit size-mass relation for dwarf galaxies from \protect\cite{ELVES2021} is shown as a solid black line, with their 1$\sigma$ uncertainties shown as black dotted lines. The colouring is consistent with Figure \ref{fig:Av_stellar_mass}.}
    \label{fig:size-mass}
\end{figure}

\begin{figure*}
    \centering
    \includegraphics[width=1\linewidth]{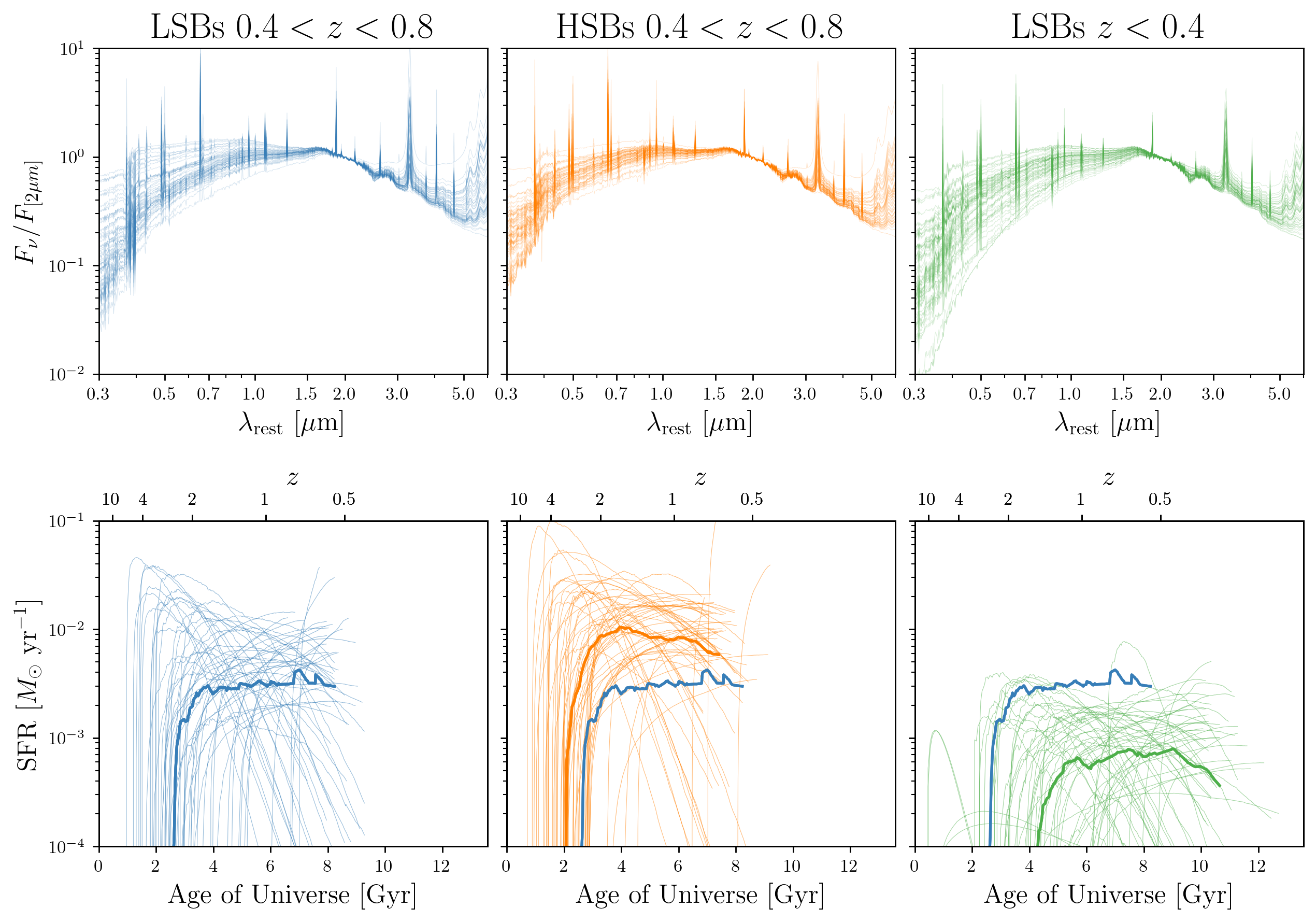}
    \caption{Posterior median spectral energy distributions (SEDs; top row) and star formation histories (SFHs; bottom row) for each sample of objects given by \texttt{bagpipes}. Each column represents a sample: the $0.4 < z < 0.8$ LSBs are on the left coloured in blue, the $0.4 < z < 0.8$ HSBs are in the middle coloured in orange, and the $z < 0.4$ LSBs are on the right coloured in green. The SEDs in the top row are fit to the photometry (see Section \ref{sec:data}) of each object by \texttt{bagpipes} and are shown from 0.3 to 6.0 microns rest frame. All SEDs are normalised to the flux at 2.0 microns rest frame, and both the relative flux and wavelengths are given as log scales. The SFHs in the bottom row show the SFR in $M_{\odot}$ yr$^{-1}$ over the age of the Universe (in Gyr on the bottom scale, in redshift on the top scale). The individual histories are given in faint lines, while the median history of the whole sample is given in a darker line. The median SFH from the higher redshift LSB sample is overplotted in the other two SFH panels for comparison.}
    \label{fig:SEDs_SFHs}
\end{figure*}

With a measure of size from the best-fit effective radius from \texttt{pyimfit} (see Section \ref{sec:selection}) and an estimate of stellar mass from \texttt{bagpipes}, we can compare our galaxy samples with previously fitted size-mass relations to see how they all fit into the general population. We use these two quantities to show a size-mass relation in Figure \ref{fig:size-mass}, and overplot the best-fit lines for dwarf ($5.5 < \log(M_*/M_{\odot}) < 8.5$) galaxies from the Exploration of Local Volume Satellites (ELVES) survey reported in \cite{ELVES2021}, along with the associated 1$\sigma$ uncertainty. 

From Figure \ref{fig:size-mass}, it is shown that most of the $0.4 < z < 0.8$ HSB sample lies along the ELVES fit within uncertainty, while most of the $0.4 < z < 0.8$ LSB sample lies above the fit with larger radii at similar stellar mass. The LSBs being more extended than the HSBs at fixed stellar mass is an effect of our selection process, as the HSBs were selected based off of having similar total magnitude but higher surface brightness, which results in the HSB sample being more compact than the LSB sample. What Figure \ref{fig:size-mass} reveals is that the higher redshift LSBs are outliers in the size-mass relation for local dwarf galaxies, whereas the majority of the HSB sample are not outliers. This supports the idea of LSBs having higher spin parameters early in their evolution \citep{perez_montano2022}, leading to extended morphologies that result in low surface brightness. Roughly half of the $z < 0.4$ LSB sample lies within the uncertainty of the ELVES fit, while the other half are extended outliers like the $0.4 < z < 0.8$ LSBs.

\subsection{Star Formation Histories}
\label{subsec:SFHs}

The star formation histories estimated by \texttt{bagpipes} for all three samples are displayed in Figure \ref{fig:SEDs_SFHs}. Each column in the figure represents a sample, with the $0.4 < z < 0.8$ LSBs on the left, the $0.4 < z < 0.8$ HSBs in the middle, and the $z < 0.4$ LSBs on the right. In the top row, we show the posterior median SED from \texttt{bagpipes} for every object in the rest frame normalised to the flux at 2 microns rest frame. The bottom row shows the posterior median SFH for each object, with the darker line of each sample showing the median of these SFHs. Since we are interested in the $0.4 < z < 0.8$ LSB sample, we overplot the sample median SFH on the other two columns for comparison.  

We show the best-fit SED of each object from \texttt{bagpipes} using the data from the JADES catalogue (see Section \ref{sec:data}). It should be noted that, although we only show the posterior median SFH for each object, the posterior distributions of SFHs for both LSB samples are not well constrained. In particular, the difference between the 16th and 84th percentile values of the onset of star formation is on average $\sim$3.5 Gyr for the individual LSB SFHs. These individual uncertainties are not shown in Figure \ref{fig:SEDs_SFHs}, but we stress the importance of spectroscopic follow-up studies on these higher-redshift LSBs in order to constrain their SFHs.

The bottom row of Figure \ref{fig:SEDs_SFHs} shows that our higher redshift LSBs start star formation anywhere from 1 Gyr to 7 Gyr after the Big Bang ($z \sim 4$ to $z \sim 0.5$). The randomly picked HSBs at similar stellar mass and redshift also start star formation at a similar range, as shown by the sample median SFHs. While the spread of SFHs in all three samples is wide, we see that it is possible for the higher redshift LSBs to start forming stars at around same point in their lives as the HSBs. The sample median history for the HSBs plateaus at a SFR of $\sim 0.01 M_{\odot}$ yr$^{-1}$ around $z \sim 2$, while the sample median history for the higher redshift LSBs plateaus at a SFR of $\sim 0.002 M_{\odot}$ yr$^{-1}$ also at around $z \sim 2$. However, this difference could be due to the HSB sample having a higher median stellar mass (see Figure \ref{fig:Av_stellar_mass}), as opposed to having inherently less star formation. Hydrodynamical simulations on LSB objects from \cite{martin2019} indicate that LSBs and HSBs come from the same progenitors at $z > 2$, but LSB progenitors form stars more rapidly early on. Interestingly, the individual SFHs in Figure \ref{fig:SEDs_SFHs} indicate that the HSBs started forming stars earlier and more rapidly. Simulations from \cite{perez_montano2022} show that spin parameters differentiate between LSBs and HSBs at $z \sim 2$. Galaxies with higher spin parameters are more likely to be extended, which decreases their surface brightness and increases the likelihood of the galaxy evolving into a LSB, as supported by our size-mass relation in Figure \ref{fig:size-mass}.

Our primary motivation for selecting a sample of LSBs at $z_{\rm phot} < 0.4$ is to explore whether or not the higher redshift LSBs in JADES evolve into the lower redshift LSBs. By looking at the bottom left and bottom right of Figure \ref{fig:SEDs_SFHs}, we see that this is a possibility, since both panels share some similar individual histories. However, the lower redshift LSB sample has many histories that drastically differ from that of the higher-redshift LSBs. Many lower-redshift LSBs started their star formation later in the history of the Universe, as shown by the lower redshift sample median SFH starting $\sim$4 Gyr later than the higher redshift sample median SFH. It’s possible that these differences in star formation represent groups of LSBs that are harder to detect past $z \sim 0.4$, and therefore aren’t seen in our higher redshift sample.

\subsection{Star Formation Rate and Mass-Weighted Age}
\label{subsec:SFR10_tMW}
 For each object, we infer a posterior distribution of the star formation rate averaged over the last 100 Myr ($\rm SFR_{100}$), and the mass-weighted age $t_{\rm MW}$. The latter is defined by \cite{BAGPIPES_carnall2018} as 

\begin{equation}
    \label{eq:tMW}
     t_{\rm MW} = \frac{\int_{0}^{t_{\rm obs}} t \: \textup{SFR}(t) \: \textup{d}t}{\int_{0}^{t_{\rm obs}} \: \textup{SFR}(t) \: \textup{d}t},
\end{equation}
where $t_{\rm obs}$ is the time at which the object is observed, $\textup{SFR}(t)$ is measured by \texttt{bagpipes}, and $t=0$ corresponds to the beginning of the Universe. The $\rm SFR_{100}$ will tell us how much recent star formation each object has, while $t_{\rm MW}$ will tell us how old the objects are for their stellar mass. 

\begin{figure}
    \centering
    \includegraphics[width=1\linewidth]{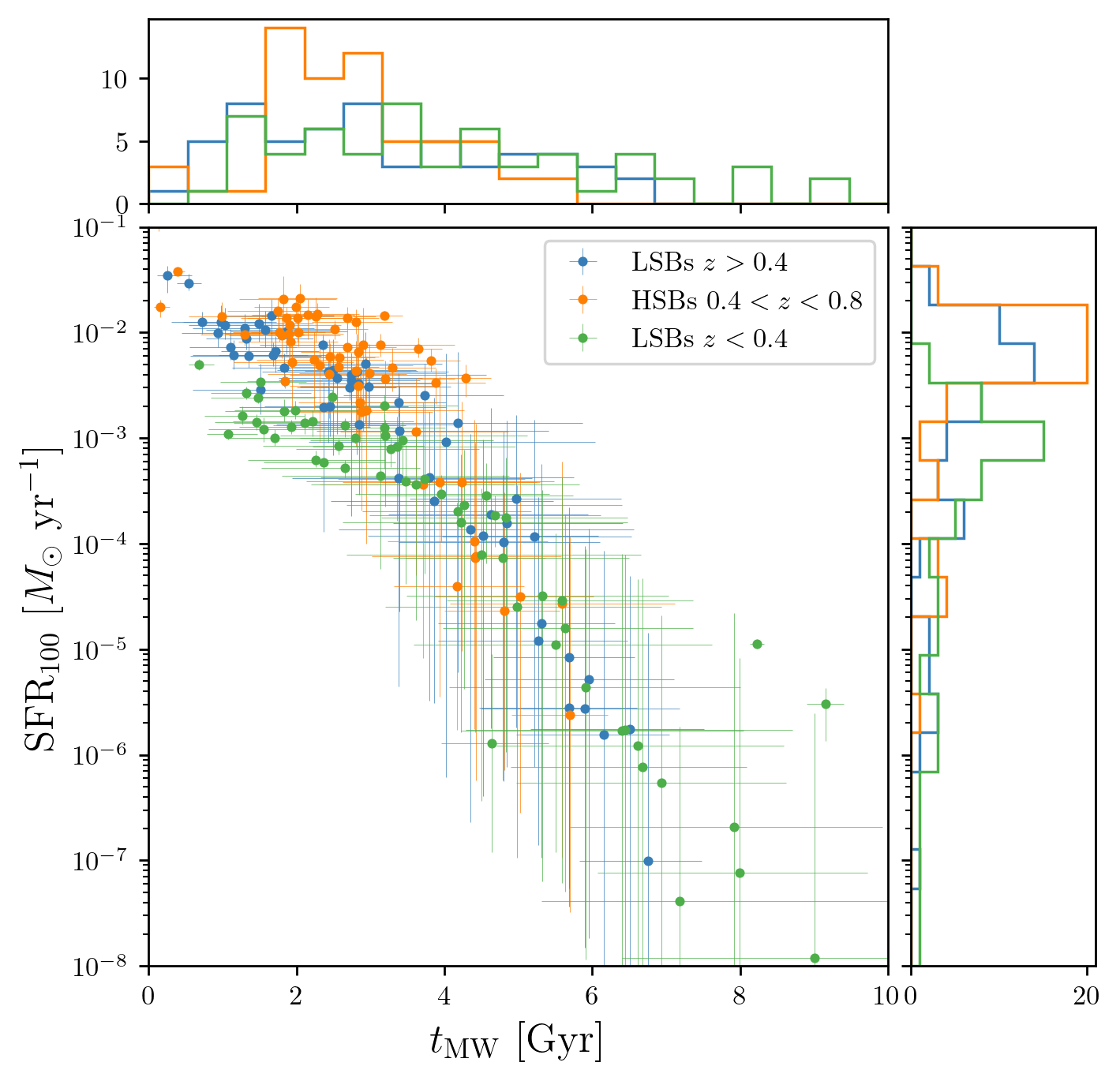}
    \caption{Star formation rate averaged over the last 100 Myr ($\rm SFR_{100}$) versus the mass weighted age ($t_{\rm MW}$, defined in Eqn \eqref{eq:tMW}) as derived from the \texttt{bagpipes} fitting. The dots represent the posterior medians, while the errorbars represent the 1$\sigma$ spread in the posterior distribution. Colouring of samples is consistent with Figures \ref{fig:Av_stellar_mass}, \ref{fig:size-mass}, and \ref{fig:SEDs_SFHs}. Histogrammes are drawn for each axis showing the distributions in both quantities.}
    \label{fig:SFR100_tMW}
\end{figure}

To explore the relationship between these two derived quantities, we plot them against each other for each object in each sample in Figure \ref{fig:SFR100_tMW} as scatter plots, where each colour represents a sample, and each dot represents the median of the posterior for an object. The dots are coloured by object type, consistent with the colouring in Figure \ref{fig:SEDs_SFHs}. 

\begin{figure*}
    \centering
    \includegraphics[width=1\linewidth]{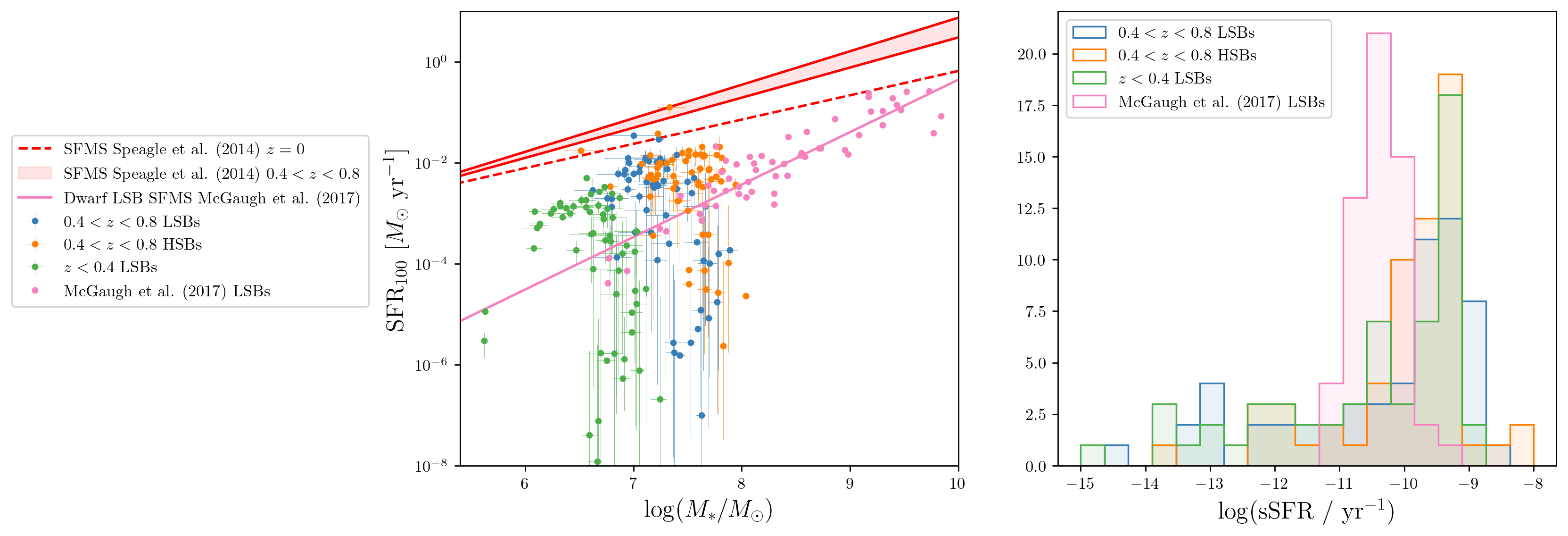}
    \caption{\textbf{Left:} SFR$_{100}$ derived from \texttt{bagpipes} for all three samples (higher redshift LSBs in blue, lower redshift LSBs in green, and HSBs in orange) plotted against the \texttt{bagpipes} stellar mass. The star forming main sequence fit for dwarf LSBs from \protect\cite{mcgaugh2017} is also shown as a pink line with the pink datapoints representing the LSBs from that study. The best-fit SFMS for all galaxies $0.4 < z < 0.8$ (shaded in red) and $z = 0$ (dotted red) are shown from \protect\cite{speagle2014}. \textbf{Right:} Distributions for the specific star formation rate (sSFR) in log scale for all four samples shown on the left panel with consistent colouring.}
    \label{fig:SFMS_sSFR}
\end{figure*}

There is not a clear difference between the $0.4 < z < 0.8$ LSBs and HSBs in terms of recent star formation (SFR$_{100}$), while the $z < 0.4$ LSBs seem to have less recent star formation than the $0.4 < z < 0.8$ LSBs. The lower redshift LSBs tend to extend to older mass-weighted ages than the $z > 0.4$ LSBs. This is supported by Figure \ref{fig:SEDs_SFHs}, which shows the low redshift LSBs having much lower star formation rate than the $0.4 < z < 0.8$ LSBs. The HSB sample does not extend past mass-weighted ages of 6 Gyr.

Figure \ref{fig:SFR100_tMW} implies that HSBs exhibit a range of mass-weighted ages between roughly 0 and 6 Gyr. High-redshift LSBs exhibit a similar range between 0 and 6 Gyr, while their lower redshift counterparts exhibit a range between 1 and 9 Gyr. Thus, the higher-redshift LSBs exhibit a much wider range in mass-weighted ages when compared to HSBs and some of the lower redshift LSBs exhibit larger mass-weighted ages than the higher-redshift LSBs.

From Figure \ref{fig:SFR100_tMW}, we can further support the conclusion that both $0.4 < z < 0.8$ samples have similar star formation histories, but the lower redshift LSBs have lower levels of star formation on average, which is all in agreement with Figure \ref{fig:SEDs_SFHs}. The LSB samples can have varying values of mass-weighted age that can be similar to that of HSBs. 

In Figure \ref{fig:SFMS_sSFR}, we report the relation between star formation rate (SFR$_{100}$) and stellar mass (left panel) both derived from \texttt{bagpipes} for all three samples of objects. We compare this with the star forming main sequence (SFMS) of dwarf LSBs found in \citep{mcgaugh2017} and show the best-fit general SFMS from \cite{speagle2014} given by 

\begin{equation}
    \textup{SFR}(M_*, t) = (0.84 - 0.026t)\log M_* - (6.51 - 0.11t),
\end{equation}
where $t$ is the age of the Universe in Gyr. We show this relation in the left panel of Figure \ref{fig:SFMS_sSFR} for the redshift range of $0.4 < z < 0.8$ (red shaded area) to reflect the range of the higher redshift LSB and HSB samples, as well as for $z=0$ (red dotted line) for the lower redshift LSB sample.

All samples seem to occupy a parameter space well below the general best-fit main sequence at their stellar mass while the dwarf LSB SFMS fits the samples well. The LSBs studied in \cite{mcgaugh2017} seem to occupy a space below the general main sequence but at higher stellar mass. This demonstrates how our study has found dwarf galaxies at lower masses and star formation rates than previous studies focusing on the SFMS of LSBs. 

In the region of Figure \ref{fig:SFMS_sSFR} where the $0.4 < z < 0.8$ LSB and HSB samples overlap in stellar mass ($7 \lesssim \log(M_* / M_{\odot}) \lesssim 8$), there is very little difference in their star formation rates, with both samples even having a similar amount of quenched galaxies in that region. This is consistent with the findings of \cite{kado_fong2022}, wherein both ultra-diffuse dwarf galaxies and "normal" dwarf galaxies have similar SFRs at fixed stellar mass. The defining characteristic of our $0.4 < z < 0.8$ LSBs, when compared to HSB dwarf galaxies at similar redshift, is their extended nature, as illustrated by Figure \ref{fig:size-mass}. Larger physical size with similar star formation history results in lower SFR density and lower surface brightness.

\section{Discussion and Conclusions}
\label{sec:conclusion}

We present a method for finding low surface brightness objects (LSBs) in the GOODS-S field using JADES data. Using the resolution and depth of the survey, as well as photometric redshifts, we have found a sample of redshift $0.4 < z < 0.8$ galaxies that are faint and extended. Using Python programmes \texttt{pyimfit} and \texttt{pysersic}, we measure S\'{e}rsic surface brightness profiles of the objects in our catalog, and select a sample of LSBs with $R_{\rm eff} > 0.18$ arcsec and $\bar{\mu}_{\rm eff} > 24$ mag arcsec$^{-2}$ in the F200W band. With these criteria, we find 57 LSBs past $z_{\rm phot} > 0.4$. 

We find that LSBs are concentrated in the areas of the survey with the greatest depth (see Figure \ref{fig:LSB_spatial_dist}). With a sample of this size, it is difficult to tell whether or not physical clustering is occurring, though LSBs are expected to be common in dense environments \citep{koda2015, greco_etal2018, roman2021}. The number density of these LSBs in our survey is on the order of $10^{-3}$ Mpc$^{-3}$, though the actual number changes with measuring the density across all of GOODS-S, as opposed to just the deepest parts of the survey. This suggests that we would need surveys of greater depth and larger area to obtain a more representative number density of LSBs at higher redshift. A crude estimate of the exposure time required to detect galaxies like those in this sample at $z \sim 2$ can be obtained by looking at the surface brightness limit in the existing mosaics combined with a simple $(1+z)^{-4}$ surface brightness extrapolation. At F200W, the $5\sigma$ surface brightness limit in $0.15^{\prime\prime} \times 0.15^{\prime\prime}$ area is $m_{AB} \sim34.7$. Using the same exposure time would yield $\mathrm{SNR} \sim 1$ detections. This suggests that at least 100 times longer exposures ($ \sim 4$ million seconds) would be needed to study their properties. 
 
Most galaxies in our sample are very extended, and with a S\'{e}rsic index of $n \sim 0.65$ (see Figure \ref{fig:multi_band_params}). However, due to our choice in using the mean effective surface brightness to define our LSBs instead of the central surface brightness, there is a group of objects in our sample with HSB centres ($\mu_0 < 23$ mag arcsec$^{-2}$). We find that in this group of outliers there are objects with bright centres and LSB disks, but there are other objects where the effective radius is overestimated, leading to a disparity in $\mu_0$ and $\mu_{\rm eff}$. The LSBs seem to be relatively faint at longer wavelengths (F356W, F444W), but have somewhat normally distributed colours at lower wavelengths (see Figure \ref{fig:color_color}). Our LSB sample is diverse in colour, although with a slight gradient toward redder in F115W-F150W and bluer in F277W-F356W colours.

Using the SED-fitting programme \texttt{bagpipes}, we estimate the star formation history, stellar mass, mass-weighted age, and dust attenuation. We compare these photometrically derived quantities with a sample of high surface brightness (HSB) objects at a similar size and stellar mass range to our LSBs, and a low redshift LSB sample defined by $z < 0.4$ and $\bar{\mu}_{\rm eff} > 24$ mag arcsec$^{-2}$ in F200W. Both LSB and HSB samples at $0.4 < z < 0.8$ appear to be quenched and void of star formation (SFR$_{100} \lesssim 10^{-2}$ $M_{\odot}$ yr$^{-1}$), and exhibit similar star formation histories and rates at fixed stellar mass, while the $z < 0.4$ LSB sample exhibits less star formation on average (see Figures \ref{fig:SEDs_SFHs}, \ref{fig:SFR100_tMW}, and  \ref{fig:SFMS_sSFR}).

Our LSB sample is low in stellar mass, and in the dwarf regime of galaxies. Both comparison samples are selected to be of comparable stellar mass using F200W magnitude as a prior before \texttt{bagpipes} fitting, though the $z < 0.4$ LSB distribution extends to notably lower masses. Both LSB samples have comparable distributions of dust attenuation, with the HSB distribution having a statistically lower attenuation. This implies dust attenuation could play a role, though the majority of our LSBs have $A_V < 1$ mag (see Figure \ref{fig:Av_stellar_mass}). Therefore, dust attenuation does not appear to be a major factor in the dimming of these LSBs for most of both samples. 

Figure \ref{fig:SEDs_SFHs} implies that, at redshifts higher than $z \sim 2$, our LSBs and the HSBs could start star formation around the same time, and therefore come from the same progenitors. Hydrodynamical simulations of LSBs from \cite{martin2019} predict this, and find that the fractions of gas in LSBs greatly diverge from HSBs around $z \sim 2$. They also show that at $z < 1$, LSBs experience large decreases in the mass of gas that isn't due to star formation, and that by $z=0$, LSBs have lost most of their star forming gas. The mechanisms from these simulations that seemed to cause the evolution of LSBs also act at this intermediate redshift range, and in the simulations are identified as supernova (SN) feedback, tidal interactions, and ram-pressure stripping (in clusters), which all serve to increase the effective radius of these galaxies and decrease their star-forming gas over time. When comparing the effective radii and stellar mass of our galaxy samples with known size-mass relations for dwarf galaxies \citep{ELVES2021}, we find that the $0.4 < z < 0.8$ are extended to the point of being outliers, whereas most of the $0.4 < z < 0.8$ HSBs lie within 1$\sigma$ of the relation. With similar star formation rates and histories, our results indicate that the increase in effective radius over time from these mechanisms leads to the formation of these LSBs.

Theoretical predictions developed by \cite{diCintio2017} and \cite{chan2018} show that SN feedback (at low stellar mass) can create ultra-diffuse galaxies by fueling outflows that remove star forming gas from the galaxies. \cite{martin2019} shows that SN feedback is the main mechanism that initializes the divergence of LSBs from HSBs past $z \sim 2$, and allows other mechanisms to act more effectively on these objects. They also show that, while ram pressure stripping is effective at removing star forming gas in dense environments, tidal heating and forces act to help form LSBs in all environments. While our sample isn't large enough to investigate whether or not the quenched galaxies form in clusters, as discussed in Sections \ref{subsec:spatial_dist} and \ref{subsec:num_dens}, the additional mechanism from ram pressure stripping might also be a factor in the quenching of these LSBs.

Further spectroscopic analysis can reveal more information about LSBs at higher redshifts and reduce the inherent uncertainties present in this study from using photometric redshifts as opposed to spectroscopic redshifts. Future surveys of greater depth can allow us to look further in the past for these LSBs, and to better constrain their star formation histories to test theoretical and cosmological predictions about their evolution and origin.

\section*{Acknowledgements}
TS, MR, JMH, KH, BJ, CNAW, YS, CC, and BR are supported by NASA contract NAS5-02105 to the University of Arizona. DJE is supported as a Simons Investigator and by JWST/NIRCam contract to the University of Arizona, NAS5-02015.  Support for programme PID 3215 was provided by NASA through a grant from the Space Telescope Science Institute, which is operated by the Association of Universities for Research in Astronomy, Inc., under NASA contract NAS 5-03127. AJB acknowledges funding from the "FirstGalaxies" Advanced Grant from the European Research Council (ERC) under the European Union’s Horizon 2020 research and innovation programmeme (Grant agreement No. 789056). ECL acknowledges support of an STFC Webb Fellowship (ST/W001438/1). The research of CCW is supported by NOIRLab, which is managed by the Association of Universities for Research in Astronomy (AURA) under a cooperative agreement with the National Science Foundation.

This work is based in part on observations made with the NASA/ESA/CSA James Webb Space Telescope. The data were obtained from the Mikulski Archive for Space Telescopes at the Space Telescope Science Institute, which is operated by the Association of Universities for Research in Astronomy, Inc., under NASA contract NAS 5-03127 for JWST. These observations are associated with the program IDs listed in the data section. The authors acknowledge P. Oesch (PID 1895), C.C. Williams (PID 1963 and 2514), S. Finkelstein (2079), and T. Morishita (PID 3990) for developing their observing programs with a zero-exclusive-access period.

This material is based upon High Performance Computing (HPC) resources supported by the University of Arizona TRIF, UITS, and Research, Innovation, and Impact (RII) and maintained by the UArizona Research Technologies department. We respectfully acknowledge the University of Arizona is on the land and territories of Indigenous peoples. Today, Arizona is home to 22 federally recognized tribes, with Tucson being home to the O'odham and the Yaqui. The University strives to build sustainable relationships with sovereign Native Nations and Indigenous communities through education offerings, partnerships, and community service.

We would like to thank our referee, Dr. Jenny Greene, for her comments and suggestions on this paper.

%%%%%%%%%%%%%%%%%%%%%%%%%%%%%%%%%%%%%%%%%%%%%%%%%%
\section*{Data Availability}

The code used to perform this analysis is publicly available on GitHub at \url{https://github.com/tristen-shields/LSBs_2025}. Cutouts of any objects used in this study not included in a JADES public data release are available upon request.



\bibliographystyle{mnras}
\bibliography{main}

%\begin{thebibliography}{99}
%\end{thebibliography}

%%%%%%%%%%%%%%%%%%%%%%%%%%%%%%%%%%%%%%%%%%%%%%%%%%

%%%%%%%%%%%%%%%%% APPENDICES %%%%%%%%%%%%%%%%%%%%%

\appendix

\section{$\texttt{pyimfit}/\texttt{pysersic}$ Comparison}
\label{app:pyimfit_pysersic}

\begin{figure*}
    \centering
    \includegraphics[width=1\linewidth]{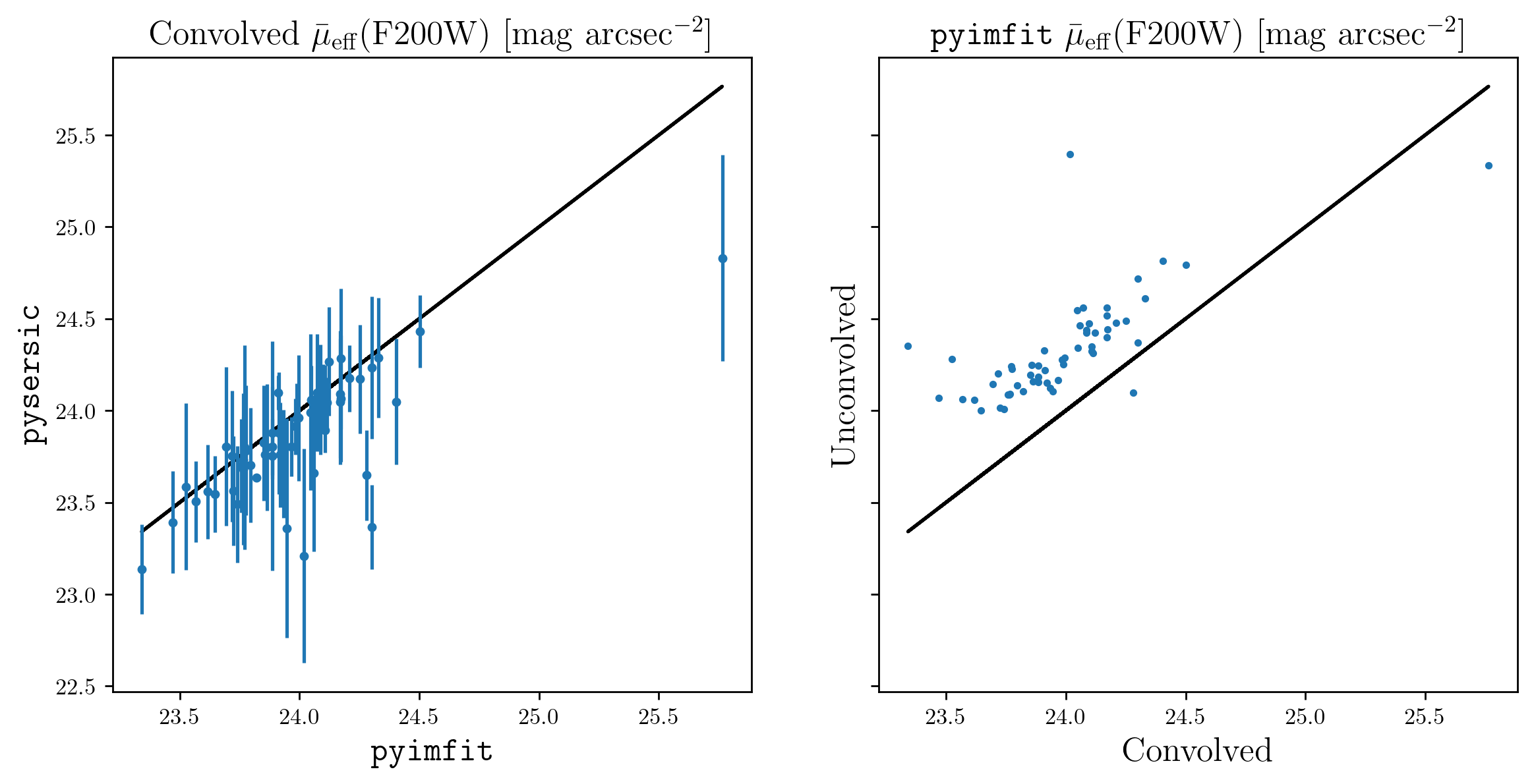}
    \caption{For the $0.4 < z_{\rm phot} < 0.8$ LSB sample, we run both \texttt{pysersic} and \texttt{pyimfit} fitting routines described in Sections \ref{sec:selection} and \ref{subsec:pysersic_profiles} respectively on the F200W band with PSF convolution and again without PSF convolution for \texttt{pyimfit}. We plot the measured mean effective surface brightness $\mu_{\rm eff}$ from both programmes against each other in the left panel, supplying errorbars for \texttt{pysersic}. On the right panel, we plot the convolved vs. unconvolved \texttt{pyimfit} mean effective surface brightnesses. The black lines in both panels represent where both brightnesses are equal.}
    \label{fig:pyimfit-pysersic-comparison}
\end{figure*}

Figure \ref{fig:pyimfit-pysersic-comparison} shows a comparison between the results of the two image fitting programmes used in this study: \texttt{pyimfit} and \texttt{pysersic}. We also directly test the effect of PSF convolution on \texttt{pyimfit} fits. PSF convolution was not used for the \texttt{pyimfit} fits described in Section \ref{sec:selection}. Here, we run convolved fits on both programmes on the $z > 0.4$ LSB sample for the F200W band and directly compare the mean effective surface brightnesses $\bar{\mu}_{\rm eff}$, the brightness parameter used for LSB selection, in the left panel of Figure \ref{fig:pyimfit-pysersic-comparison}. Errors estimated from \texttt{pysersic} are included. From the right panel of Figure \ref{fig:pyimfit-pysersic-comparison}, we can see that PSF convolution raises the brightness of the LSB \texttt{pyimfit} results. 

\section{Data for Galaxies in the LSB Sample}
In this appendix, we report data tables for each galaxy in our $0.4 < z_{\rm phot} < 0.8$ LSB sample. Table \ref{tab:pyimfit} reports the right ascension, declination, and the unconvolved F200W \texttt{pyimfit} surface brightness fitting results. Table \ref{tab:pysersic} reports the convolved multi-band \texttt{pysersic} surface brightness fitting results. Table \ref{tab:BAGPIPES} reports the derived stellar masses, dust attenuations, mass-weighted ages, and star formations averaged over the last 100 Myr. Table \ref{tab:fluxes} reports the fluxes in chosen NIRCam wide-band filters for the ``small Kron" aperture from the photometric catalogue.

\begin{table*}
    \caption{Observable quantities and \texttt{pyimfit} results (see Section \ref{sec:selection}) for the $z > 0.4$ LSB sample. For each object ID, we report the RA (right ascension) and declination from the photometric catalogue described in Section \ref{sec:data}. The photometric redshift used from \texttt{eazy-py} is shown in the $z_{\rm phot}$ column. From \texttt{pyimfit} we report the mean effective surface brightness measured in F200W ($\bar{\mu}_{\rm eff}$(F200W)) in magnitudes per square arcsecond, the central surface brightness measured in F200W ($\mu_0$(F200W)) in magnitudes per square arcsecond, the effective radius $R_{\rm eff}$ in arcsec, and the S\'{e}rsic index $n$.}
    \label{tab:pyimfit}
    \centering
    \begin{tabular}{c|c|c|c|c|c|c}
        & JADES ID & $z_{\rm phot}$ & \makecell{\texttt{pyimfit} \\ $\bar{\mu}_{\rm eff}$(F200W) \\ mag arcsec$^{-2}$} & \makecell{\texttt{pyimfit} \\ $\mu_0$(F200W) \\ mag arcsec$^{-2}$} & \makecell{\texttt{pyimfit} \\ $R_{\rm eff}$ \\ arcsec} & \makecell{\texttt{pyimfit} \\ $n$} \\
        \hline 
JADES-GS-53.04873--27.90275 & 5281 & 0.57 & 24.72 & 23.73 & 0.19 & 0.92 \\ 
JADES-GS-53.03687--27.87839 & 29405 & 0.83 & 24.24 & 22.46 & 0.21 & 1.37 \\ 
JADES-GS-53.05435--27.87539 & 34150 & 0.70 & 24.24 & 22.90 & 0.18 & 1.13 \\ 
JADES-GS-53.12571--27.85182 & 69546 & 0.59 & 24.06 & 21.88 & 0.31 & 1.59 \\ 
JADES-GS-53.15494--27.80422 & 109405 & 0.43 & 24.48 & 23.30 & 0.28 & 1.03 \\ 
JADES-GS-53.12658--27.79770 & 113557 & 0.74 & 24.35 & 22.61 & 0.19 & 1.35 \\ 
JADES-GS-53.17576--27.79028 & 118072 & 0.42 & 24.47 & 23.39 & 0.19 & 0.97 \\ 
JADES-GS-53.16779--27.78889 & 119192 & 0.50 & 24.00 & 23.33 & 0.20 & 0.72 \\ 
JADES-GS-53.15466--27.77882 & 127231 & 0.62 & 24.55 & 23.99 & 0.18 & 0.64 \\ 
JADES-GS-53.21008--27.75993 & 143098 & 0.41 & 25.33 & 24.04 & 0.21 & 1.10 \\ 
JADES-GS-53.03636--27.89731 & 160858 & 0.55 & 24.10 & 21.27 & 0.26 & 1.94 \\ 
JADES-GS-53.05608--27.88765 & 164373 & 0.56 & 24.22 & 23.87 & 0.27 & 0.49 \\ 
JADES-GS-53.03624--27.87367 & 171611 & 0.78 & 24.16 & 23.06 & 0.21 & 0.99 \\ 
JADES-GS-53.06586--27.86710 & 175180 & 0.55 & 24.14 & 23.36 & 0.19 & 0.79 \\ 
JADES-GS-53.08662--27.85548 & 183481 & 0.70 & 24.01 & 22.25 & 0.19 & 1.36 \\ 
JADES-GS-53.08983--27.84115 & 189735 & 0.53 & 24.40 & 24.01 & 0.21 & 0.53 \\ 
JADES-GS-53.07777--27.83917 & 190476 & 0.55 & 24.09 & 23.39 & 0.23 & 0.74 \\ 
JADES-GS-53.16757--27.82074 & 196612 & 0.55 & 24.15 & 23.16 & 0.18 & 0.92 \\ 
JADES-GS-53.19068--27.81485 & 198340 & 0.59 & 24.23 & 23.18 & 0.21 & 0.95 \\ 
JADES-GS-53.07564--27.81459 & 198413 & 0.56 & 24.01 & 23.33 & 0.20 & 0.73 \\ 
JADES-GS-53.15474--27.81125 & 199235 & 0.55 & 24.61 & 24.16 & 0.35 & 0.57 \\ 
JADES-GS-53.11976--27.78368 & 207083 & 0.59 & 24.25 & 23.70 & 0.23 & 0.64 \\ 
JADES-GS-53.19289--27.78256 & 207340 & 0.55 & 24.32 & 23.05 & 0.34 & 1.09 \\ 
JADES-GS-53.20654--27.77870 & 208568 & 0.69 & 24.52 & 23.79 & 0.20 & 0.76 \\ 
JADES-GS-53.15625--27.77703 & 209020 & 0.48 & 24.49 & 23.91 & 0.30 & 0.66 \\ 
JADES-GS-53.14173--27.77264 & 210372 & 0.56 & 24.42 & 23.61 & 0.24 & 0.81 \\ 
JADES-GS-53.17890--27.74961 & 216996 & 0.57 & 24.29 & 23.91 & 0.20 & 0.51 \\ 
JADES-GS-53.17294--27.74677 & 217509 & 0.63 & 24.06 & 22.72 & 0.20 & 1.12 \\ 
JADES-GS-53.14745--27.74365 & 217847 & 0.47 & 24.15 & 22.86 & 0.25 & 1.10 \\ 
JADES-GS-53.09791--27.74248 & 235455 & 0.43 & 24.31 & 23.45 & 0.34 & 0.84 \\ 
JADES-GS-53.17355--27.78914 & 285675 & 0.54 & 24.81 & 23.68 & 0.18 & 1.01 \\ 
JADES-GS-53.11639--27.88056 & 301713 & 0.48 & 24.07 & 22.24 & 0.18 & 1.40 \\ 
JADES-GS-53.10587--27.77242 & 313445 & 0.59 & 24.14 & 23.39 & 0.25 & 0.77 \\ 
JADES-GS-53.11345--27.75675 & 315445 & 0.50 & 24.11 & 22.30 & 0.20 & 1.39 \\ 
JADES-GS-53.13969--27.89341 & 318555 & 0.72 & 24.28 & 22.41 & 0.31 & 1.42 \\ 
JADES-GS-53.25539--27.87473 & 415856 & 0.61 & 24.12 & 22.69 & 0.21 & 1.18 \\ 
JADES-GS-53.26882--27.87448 & 415974 & 0.41 & 24.79 & 23.86 & 0.22 & 0.88 \\ 
JADES-GS-53.25091--27.87330 & 416526 & 0.53 & 24.44 & 23.73 & 0.22 & 0.74 \\ 
JADES-GS-53.26900--27.86701 & 419511 & 0.41 & 24.42 & 23.53 & 0.19 & 0.86 \\ 
JADES-GS-53.24986--27.86703 & 419512 & 0.60 & 24.09 & 22.93 & 0.25 & 1.02 \\ 
JADES-GS-53.26452--27.86380 & 420939 & 0.56 & 24.25 & 23.40 & 0.29 & 0.84 \\ 
JADES-GS-53.24263--27.85051 & 427953 & 0.51 & 24.56 & 23.25 & 0.25 & 1.11 \\ 
JADES-GS-53.25589--27.82759 & 445205 & 0.51 & 24.28 & 23.22 & 0.23 & 0.96 \\ 
JADES-GS-53.02777--27.82618 & 446697 & 0.44 & 24.19 & 23.14 & 0.19 & 0.96 \\ 
JADES-GS-53.01088--27.81897 & 453343 & 0.58 & 24.20 & 23.43 & 0.25 & 0.78 \\ 
JADES-GS-53.00647--27.81514 & 456812 & 0.50 & 24.18 & 22.88 & 0.26 & 1.11 \\ 
JADES-GS-52.98825--27.80582 & 463900 & 0.44 & 24.37 & 22.21 & 0.23 & 1.58 \\ 
JADES-GS-53.07405--27.71603 & 493737 & 0.45 & 24.33 & 22.00 & 0.31 & 1.67 \\ 
JADES-GS-53.27421--27.86730 & 514992 & 0.48 & 25.39 & 20.89 & 0.23 & 2.79 \\ 
JADES-GS-53.26607--27.84964 & 520754 & 0.55 & 24.34 & 23.49 & 0.25 & 0.84 \\ 
JADES-GS-53.03591--27.83111 & 527971 & 0.60 & 24.10 & 23.77 & 0.20 & 0.48 \\ 
JADES-GS-53.00424--27.82547 & 530989 & 0.41 & 24.17 & 22.44 & 0.33 & 1.35 \\ 
JADES-GS-53.00282--27.82417 & 531703 & 0.58 & 24.44 & 22.73 & 0.29 & 1.34 \\ 
JADES-GS-53.02628--27.82051 & 533870 & 0.49 & 24.46 & 23.19 & 0.19 & 1.08 \\ 
JADES-GS-53.24369--27.80327 & 541593 & 0.49 & 24.35 & 23.56 & 0.34 & 0.79 \\ 
JADES-GS-52.97845--27.79245 & 545495 & 0.45 & 24.56 & 23.48 & 0.20 & 0.98 \\ 
JADES-GS-53.00096--27.78652 & 547212 & 0.57 & 24.09 & 22.49 & 0.20 & 1.28 \\ 
    \end{tabular}
    \label{tab:pyimfit}
\end{table*}

\newcolumntype{C}[1]{>{\centering\arraybackslash}p{#1}}
\newcolumntype{Y}{>{\centering\arraybackslash}X}
\begin{table*}
    \caption{Results from best fits from \texttt{pysersic} (see \ref{subsec:pysersic_profiles} in four different bands: F150W, F200W, F277W, and F356W. For each band, we show the mean effective surface brightness $\bar{\mu}_{\rm eff}$ in magnitudes per square arcsecond, the central surface brightness $\mu_0$ measured in magnitudes per square arcsecond, the effective radius $R_{\rm eff}$ in arcsec, and the S\'{e}rsic index $n$. For each of these quantities, there are four separate columns listing the measurements in each band. The corresponding object IDs are in the leftmost column.}
    \centering
    \begin{tabularx}{\textwidth}{c|X|X|X|X|X|X|X|X|X|X|X|X|X|X|X|X}
         & \multicolumn{4}{c|}{$\bar{\mu}_{\rm eff}$ [mag arcsec$^{-2}$]} & \multicolumn{4}{c|}{$\mu_0$ [mag arcsec$^{-2}$]} & \multicolumn{4}{c|}{$R_{\rm eff}$ [arcsec]} & \multicolumn{4}{c}{$n$} \\ \hline
        ID & \hspace*{-0.8em} F150W &\hspace*{-0.8em} F200W &\hspace*{-0.8em} F277W &\hspace*{-0.8em} F356W &\hspace*{-0.8em} F150W &\hspace*{-0.8em} F200W &\hspace*{-0.8em} F277W &\hspace*{-0.8em} F356W &\hspace*{-0.8em} F150W &\hspace*{-0.8em} F200W &\hspace*{-0.8em} F277W &\hspace*{-0.8em} F356W &\hspace*{-0.8em} F150W &\hspace*{-0.8em} F200W &\hspace*{-0.8em} F277W &\hspace*{-0.8em} F356W \\ \hline
        5281 & 24.54 & 24.23 & 24.28 & 24.35 & 21.50 & 22.81 & 23.47 & 23.10 & 0.16 & 0.16 & 0.14 & 0.17 & 2.04 & 1.17 & 0.81 & 1.08 \\ 
        29405 & 24.18 & 23.75 & 24.35 & 24.93 & 19.74 & 20.91 & 19.66 & 20.68 & 0.37 & 0.20 & 0.30 & 0.38 & 2.76 & 1.94 & 2.89 & 2.66 \\ 
        34150 & 23.61 & 23.80 & 23.95 & 23.92 & 21.42 & 21.64 & 21.20 & 22.71 & 0.13 & 0.17 & 0.23 & 0.21 & 1.60 & 1.58 & 1.89 & 1.05 \\ 
        69546 & 23.48 & 23.50 & 23.27 & 23.26 & 21.13 & 21.78 & 21.52 & 22.38 & 0.35 & 0.33 & 0.39 & 0.29 & 1.68 & 1.34 & 1.36 & 0.85 \\ 
        109405 & 24.09 & 24.18 & 24.16 & 24.81 & 22.82 & 22.68 & 22.13 & 21.94 & 0.23 & 0.25 & 0.24 & 0.30 & 1.08 & 1.22 & 1.51 & 1.96 \\ 
        113557 & 23.45 & 23.14 & 23.32 & 23.26 & 22.46 & 21.71 & 22.75 & 22.69 & 0.18 & 0.19 & 0.14 & 0.18 & 0.92 & 1.18 & 0.65 & 0.65 \\ 
        118072 & 24.29 & 24.07 & 24.28 & 24.63 & 22.32 & 22.81 & 22.52 & 23.57 & 0.23 & 0.20 & 0.22 & 0.22 & 1.48 & 1.08 & 1.36 & 0.96 \\ 
        119192 & 23.60 & 23.55 & 23.61 & 23.70 & 22.73 & 22.74 & 22.38 & 22.86 & 0.17 & 0.17 & 0.21 & 0.18 & 0.84 & 0.81 & 1.06 & 0.83 \\ 
        127231 & 24.40 & 23.99 & 24.56 & 24.78 & 22.89 & 22.98 & 22.61 & 21.30 & 0.11 & 0.16 & 0.18 & 0.17 & 1.22 & 0.93 & 1.47 & 2.27 \\ 
        143098 & 24.62 & 24.83 & 24.62 & 25.13 & 23.91 & 22.90 & 24.06 & 24.43 & 0.15 & 0.18 & 0.21 & 0.17 & 0.75 & 1.45 & 0.65 & 0.74 \\ 
        160858 & 23.33 & 23.65 & 23.08 & 23.76 & 19.96 & 19.23 & 19.89 & 18.94 & 0.17 & 0.23 & 0.16 & 0.20 & 2.22 & 2.75 & 2.12 & 2.95 \\ 
        164373 & 23.84 & 23.88 & 24.54 & 24.49 & 23.27 & 23.30 & 22.11 & 23.93 & 0.25 & 0.24 & 0.40 & 0.39 & 0.65 & 0.66 & 1.73 & 0.65 \\ 
        171611 & 24.02 & 23.80 & 23.60 & 23.89 & 23.19 & 22.23 & 22.76 & 22.44 & 0.20 & 0.20 & 0.17 & 0.23 & 0.82 & 1.25 & 0.83 & 1.19 \\ 
        175180 & 24.13 & 23.70 & 24.04 & 24.46 & 22.26 & 22.92 & 22.44 & 22.10 & 0.26 & 0.19 & 0.27 & 0.26 & 1.42 & 0.79 & 1.27 & 1.69 \\ 
        183481 & 23.24 & 23.49 & 22.98 & 23.51 & 22.46 & 20.83 & 21.23 & 21.35 & 0.14 & 0.17 & 0.13 & 0.20 & 0.79 & 1.85 & 1.36 & 1.58 \\ 
        189735 & 24.38 & 24.28 & 24.20 & 24.50 & 23.49 & 23.42 & 23.63 & 23.94 & 0.28 & 0.23 & 0.22 & 0.21 & 0.86 & 0.84 & 0.65 & 0.65 \\ 
        190476 & 23.69 & 23.70 & 23.58 & 23.89 & 22.62 & 22.98 & 22.72 & 23.21 & 0.22 & 0.23 & 0.23 & 0.20 & 0.97 & 0.75 & 0.84 & 0.72 \\ 
        196612 & 23.65 & 23.76 & 24.34 & 24.54 & 22.59 & 22.25 & 20.95 & 20.81 & 0.14 & 0.17 & 0.26 & 0.25 & 0.96 & 1.22 & 2.23 & 2.40 \\ 
        198340 & 23.91 & 23.78 & 24.24 & 24.78 & 21.84 & 22.29 & 22.41 & 22.98 & 0.20 & 0.18 & 0.22 & 0.22 & 1.54 & 1.21 & 1.40 & 1.38 \\ 
        198413 & 23.65 & 23.56 & 23.42 & 24.15 & 23.09 & 22.91 & 22.85 & 22.84 & 0.20 & 0.19 & 0.20 & 0.21 & 0.65 & 0.71 & 0.65 & 1.11 \\ 
        199235 & 24.59 & 24.29 & 24.63 & 25.03 & 23.65 & 23.71 & 23.37 & 24.01 & 0.42 & 0.32 & 0.44 & 0.44 & 0.89 & 0.66 & 1.08 & 0.94 \\ 
        207083 & 23.70 & 23.76 & 23.91 & 24.20 & 23.14 & 23.20 & 23.34 & 23.63 & 0.21 & 0.22 & 0.21 & 0.22 & 0.65 & 0.65 & 0.65 & 0.65 \\ 
        207340 & 23.93 & 23.89 & 23.78 & 23.95 & 22.57 & 22.69 & 23.00 & 23.18 & 0.41 & 0.38 & 0.33 & 0.30 & 1.13 & 1.04 & 0.79 & 0.78 \\ 
        208568 & 23.96 & 24.09 & 24.34 & 24.48 & 22.63 & 23.17 & 22.80 & 23.92 & 0.17 & 0.17 & 0.19 & 0.18 & 1.12 & 0.88 & 1.23 & 0.65 \\ 
        209020 & 24.36 & 24.17 & 24.39 & 24.68 & 23.75 & 23.58 & 23.60 & 24.10 & 0.26 & 0.26 & 0.27 & 0.24 & 0.68 & 0.67 & 0.79 & 0.66 \\ 
        210372 & 24.02 & 24.27 & 24.19 & 24.96 & 23.45 & 23.02 & 23.44 & 23.48 & 0.20 & 0.26 & 0.25 & 0.37 & 0.65 & 1.07 & 0.77 & 1.20 \\ 
        216996 & 23.94 & 23.96 & 24.36 & 24.33 & 23.05 & 23.39 & 23.80 & 23.07 & 0.18 & 0.18 & 0.22 & 0.23 & 0.86 & 0.66 & 0.65 & 1.08 \\ 
        217509 & 23.81 & 23.56 & 23.68 & 23.73 & 21.36 & 22.27 & 22.31 & 22.27 & 0.25 & 0.24 & 0.28 & 0.28 & 1.74 & 1.09 & 1.14 & 1.19 \\ 
        217847 & 23.83 & 23.80 & 23.83 & 24.21 & 22.26 & 22.20 & 22.13 & 22.59 & 0.24 & 0.22 & 0.25 & 0.26 & 1.25 & 1.27 & 1.33 & 1.28 \\ 
        235455 & 24.21 & 24.04 & 24.05 & 24.46 & 23.20 & 23.20 & 22.97 & 23.05 & 0.33 & 0.33 & 0.35 & 0.34 & 0.93 & 0.83 & 0.98 & 1.16 \\ 
        285675 & 24.02 & 24.05 & 24.46 & 24.63 & 23.06 & 23.02 & 22.91 & 22.94 & 0.23 & 0.20 & 0.29 & 0.29 & 0.90 & 0.94 & 1.25 & 1.32 \\ 
        301713 & 23.72 & 23.39 & 23.29 & 23.38 & 21.86 & 21.08 & 21.62 & 22.47 & 0.22 & 0.18 & 0.16 & 0.16 & 1.42 & 1.66 & 1.31 & 0.87 \\ 
        313445 & 23.85 & 23.80 & 23.63 & 24.13 & 22.81 & 22.53 & 23.07 & 21.91 & 0.21 & 0.22 & 0.28 & 0.31 & 0.95 & 1.09 & 0.65 & 1.61 \\ 
        315445 & 23.18 & 23.36 & 23.25 & 23.97 & 21.25 & 20.77 & 21.12 & 21.71 & 0.21 & 0.16 & 0.18 & 0.21 & 1.46 & 1.81 & 1.56 & 1.63 \\ 
        318555 & 23.58 & 23.59 & 23.51 & 23.34 & 21.83 & 21.92 & 22.94 & 22.50 & 0.27 & 0.28 & 0.21 & 0.26 & 1.36 & 1.31 & 0.65 & 0.83 \\ 
        415856 & 23.72 & 23.71 & 24.00 & 24.49 & 20.81 & 21.41 & 19.24 & 20.07 & 0.17 & 0.19 & 0.23 & 0.24 & 1.98 & 1.66 & 2.92 & 2.75 \\ 
        415974 & 24.31 & 24.43 & 24.20 & 24.76 & 23.38 & 23.28 & 23.54 & 24.02 & 0.19 & 0.21 & 0.17 & 0.19 & 0.89 & 1.02 & 0.71 & 0.77 \\ 
        416526 & 23.93 & 24.06 & 24.20 & 24.82 & 23.36 & 23.20 & 23.53 & 23.54 & 0.17 & 0.19 & 0.21 & 0.25 & 0.65 & 0.84 & 0.72 & 1.09 \\ 
        419511 & 24.02 & 23.97 & 24.11 & 24.40 & 23.39 & 23.01 & 22.74 & 23.83 & 0.19 & 0.19 & 0.21 & 0.17 & 0.70 & 0.90 & 1.15 & 0.65 \\ 
        419512 & 23.74 & 23.73 & 23.62 & 23.84 & 22.73 & 22.34 & 22.51 & 22.85 & 0.22 & 0.22 & 0.22 & 0.21 & 0.93 & 1.16 & 0.99 & 0.92 \\ 
        420939 & 24.09 & 23.96 & 24.04 & 24.38 & 23.32 & 23.09 & 23.22 & 23.81 & 0.26 & 0.28 & 0.30 & 0.25 & 0.78 & 0.85 & 0.81 & 0.65 \\ 
        427953 & 24.06 & 24.10 & 24.72 & 23.85 & 23.17 & 22.37 & 22.49 & 23.05 & 0.23 & 0.24 & 0.44 & 0.20 & 0.86 & 1.35 & 1.62 & 0.80 \\ 
        445205 & 23.99 & 23.91 & 23.84 & 24.24 & 22.38 & 22.68 & 22.53 & 23.13 & 0.21 & 0.25 & 0.23 & 0.27 & 1.28 & 1.07 & 1.11 & 0.99 \\ 
        446697 & 24.20 & 23.82 & 23.53 & 23.68 & 22.65 & 22.30 & 22.87 & 23.03 & 0.22 & 0.18 & 0.14 & 0.11 & 1.25 & 1.23 & 0.71 & 0.71 \\ 
        453343 & 23.84 & 23.75 & 24.82 & 25.25 & 23.26 & 22.61 & 24.26 & 23.39 & 0.22 & 0.24 & 0.34 & 0.35 & 0.67 & 1.01 & 0.65 & 1.42 \\ 
        456812 & 23.93 & 23.88 & 23.75 & 23.88 & 22.20 & 22.19 & 22.16 & 22.76 & 0.29 & 0.24 & 0.27 & 0.20 & 1.34 & 1.32 & 1.27 & 1.00 \\ 
        463900 & 23.25 & 23.37 & 23.02 & 23.25 & 21.42 & 21.27 & 21.80 & 22.67 & 0.20 & 0.22 & 0.14 & 0.12 & 1.40 & 1.55 & 1.06 & 0.66 \\ 
        493737 & 24.11 & 24.10 & 24.08 & 24.31 & 21.86 & 20.58 & 21.73 & 22.69 & 0.30 & 0.32 & 0.32 & 0.32 & 1.63 & 2.29 & 1.68 & 1.29 \\ 
        514992 & 22.01 & 23.21 & 25.00 & 23.95 & 18.97 & 19.25 & 21.42 & 19.83 & 0.07 & 0.13 & 0.25 & 0.16 & 2.04 & 2.52 & 2.32 & 2.60 \\ 
        520754 & 24.41 & 24.06 & 24.20 & 24.70 & 23.26 & 23.06 & 23.10 & 23.41 & 0.35 & 0.23 & 0.25 & 0.35 & 1.02 & 0.92 & 0.98 & 1.10 \\ 
        527971 & 23.72 & 23.63 & 23.43 & 23.93 & 23.16 & 23.07 & 22.87 & 23.37 & 0.21 & 0.19 & 0.22 & 0.23 & 0.65 & 0.65 & 0.65 & 0.65 \\ 
        530989 & 23.71 & 23.80 & 23.96 & 24.05 & 21.65 & 21.76 & 21.64 & 22.31 & 0.26 & 0.30 & 0.34 & 0.28 & 1.53 & 1.52 & 1.66 & 1.35 \\ 
        531703 & 24.03 & 24.07 & 23.65 & 23.56 & 21.70 & 21.83 & 22.21 & 22.99 & 0.29 & 0.26 & 0.19 & 0.14 & 1.67 & 1.62 & 1.18 & 0.65 \\ 
        533870 & 23.20 & 23.66 & 23.46 & 24.07 & 22.04 & 22.49 & 22.71 & 22.90 & 0.14 & 0.14 & 0.14 & 0.20 & 1.02 & 1.02 & 0.77 & 1.03 \\ 
        541593 & 24.18 & 24.08 & 24.11 & 24.37 & 23.03 & 23.29 & 23.33 & 23.74 & 0.33 & 0.30 & 0.31 & 0.27 & 1.01 & 0.80 & 0.79 & 0.69 \\ 
        545495 & 24.43 & 24.05 & 24.08 & 24.30 & 21.75 & 22.65 & 22.90 & 22.78 & 0.21 & 0.17 & 0.19 & 0.17 & 1.86 & 1.16 & 1.03 & 1.23 \\ 
        547212 & 23.92 & 23.68 & 23.57 & 23.65 & 21.06 & 21.17 & 21.43 & 22.88 & 0.20 & 0.18 & 0.24 & 0.14 & 1.95 & 1.77 & 1.57 & 0.78 \\ 

    \end{tabularx}
    \label{tab:pysersic}
\end{table*}

\begin{table*}
    \caption{Results from \texttt{bagpipes} for the $z > 0.4$ LSB sample, shown in Figures \ref{fig:SFR100_tMW} and \ref{fig:Av_stellar_mass}. We report the star formation rate averaged over the last 100 Myr from the object's epoch of observation SFR$_{100}$ in solar masses per year, stellar mass as log10 of the stellar mass divided by solar mass $\log(M_* / M_{\odot})$, and V-band dust attenuation $A_V$ in AB magnitudes.}
    \centering
    \begin{tabular}{c|c|c|c|c}
       ID  & SFR$_{100}$ [$M_{\odot}$ yr$^{-1}$] & $\log(M_* / M_{\odot})$ & $t_{\rm MW}$ [Gyr] & $A_V$ [AB mag] \\
       \hline
5281 & 0.0020 +0.0014/-0.0018 & 6.80 +0.22/-0.19 & 2.36 +1.66/-1.14 & 0.52 +0.33/-0.32 \\ 
29405 & 0.0127 +0.0054/-0.0035 & 7.11 +0.10/-0.15 & 0.98 +0.74/-0.51 & 0.23 +0.17/-0.14 \\ 
34150 & 0.0004 +0.0037/-0.0004 & 7.16 +0.13/-0.19 & 3.38 +1.81/-1.44 & 0.63 +0.34/-0.35 \\ 
69546 & 0.0002 +0.0017/-0.0002 & 7.89 +0.07/-0.12 & 4.63 +1.32/-1.38 & 0.74 +0.37/-0.37 \\ 
109405 & 0.0000 +0.0001/-0.0000 & 7.59 +0.06/-0.09 & 5.96 +1.15/-1.41 & 0.82 +0.10/-0.18 \\ 
113557 & 0.0067 +0.0026/-0.0025 & 7.05 +0.07/-0.10 & 1.72 +0.59/-0.74 & 0.51 +0.16/-0.18 \\ 
118072 & 0.0022 +0.0005/-0.0006 & 7.05 +0.07/-0.08 & 3.38 +0.64/-0.85 & 0.07 +0.05/-0.04 \\ 
119192 & 0.0000 +0.0001/-0.0000 & 7.43 +0.04/-0.06 & 6.16 +0.89/-1.19 & 0.63 +0.09/-0.17 \\ 
127231 & 0.0029 +0.0021/-0.0012 & 6.62 +0.13/-0.17 & 1.51 +0.97/-0.92 & 1.00 +0.29/-0.39 \\ 
143098 & 0.0001 +0.0010/-0.0001 & 6.84 +0.14/-0.24 & 4.36 +2.21/-1.79 & 0.71 +0.28/-0.41 \\ 
160858 & 0.0043 +0.0055/-0.0036 & 7.50 +0.15/-0.17 & 2.81 +1.24/-0.94 & 0.66 +0.23/-0.34 \\ 
164373 & 0.0110 +0.0041/-0.0030 & 7.14 +0.17/-0.23 & 1.30 +0.94/-0.71 & 0.62 +0.17/-0.31 \\ 
171611 & 0.0144 +0.0063/-0.0076 & 7.40 +0.09/-0.09 & 1.66 +0.84/-0.68 & 1.23 +0.20/-0.26 \\ 
175180 & 0.0035 +0.0017/-0.0015 & 7.23 +0.08/-0.08 & 2.74 +0.93/-0.98 & 0.07 +0.09/-0.05 \\ 
183481 & 0.0044 +0.0043/-0.0037 & 7.28 +0.14/-0.12 & 2.49 +1.13/-0.76 & 0.38 +0.36/-0.27 \\ 
189735 & 0.0042 +0.0019/-0.0033 & 7.12 +0.20/-0.13 & 2.43 +1.12/-0.89 & 0.17 +0.17/-0.12 \\ 
190476 & 0.0051 +0.0050/-0.0037 & 7.56 +0.10/-0.09 & 2.93 +1.04/-0.99 & 0.67 +0.13/-0.14 \\ 
196612 & 0.0033 +0.0025/-0.0022 & 7.19 +0.09/-0.09 & 2.83 +1.05/-0.81 & 0.98 +0.16/-0.17 \\ 
198340 & 0.0073 +0.0035/-0.0018 & 6.92 +0.10/-0.15 & 1.11 +0.89/-0.56 & 0.16 +0.15/-0.11 \\ 
198413 & 0.0040 +0.0024/-0.0038 & 7.21 +0.27/-0.11 & 2.74 +1.83/-0.83 & 0.21 +0.24/-0.15 \\ 
199235 & 0.0000 +0.0001/-0.0000 & 7.53 +0.05/-0.07 & 5.69 +0.92/-1.22 & 0.46 +0.17/-0.32 \\ 
207083 & 0.0087 +0.0024/-0.0018 & 7.06 +0.14/-0.10 & 1.31 +0.91/-0.54 & 0.04 +0.05/-0.03 \\ 
207340 & 0.0293 +0.0068/-0.0041 & 7.23 +0.05/-0.05 & 0.54 +0.17/-0.16 & 0.13 +0.10/-0.08 \\ 
208568 & 0.0047 +0.0016/-0.0016 & 6.96 +0.08/-0.09 & 1.83 +0.69/-0.71 & 0.18 +0.17/-0.12 \\ 
209020 & 0.0000 +0.0000/-0.0000 & 7.63 +0.03/-0.05 & 6.76 +0.72/-0.94 & 0.38 +0.11/-0.17 \\ 
210372 & 0.0031 +0.0019/-0.0018 & 7.40 +0.11/-0.12 & 2.98 +1.11/-0.94 & 0.17 +0.14/-0.12 \\ 
216996 & 0.0009 +0.0054/-0.0009 & 7.30 +0.14/-0.18 & 4.02 +2.02/-1.86 & 0.34 +0.27/-0.18 \\ 
217509 & 0.0349 +0.0081/-0.0113 & 7.00 +0.21/-0.24 & 0.25 +0.34/-0.14 & 1.51 +0.11/-0.15 \\ 
217847 & 0.0001 +0.0014/-0.0001 & 7.64 +0.07/-0.10 & 5.22 +1.31/-1.25 & 0.13 +0.18/-0.10 \\ 
235455 & 0.0002 +0.0013/-0.0002 & 7.79 +0.08/-0.13 & 4.85 +1.57/-1.54 & 0.15 +0.24/-0.11 \\ 
285675 & 0.0061 +0.0023/-0.0016 & 6.86 +0.12/-0.11 & 1.15 +0.99/-0.56 & 1.21 +0.15/-0.15 \\ 
301713 & 0.0125 +0.0032/-0.0028 & 6.95 +0.26/-0.22 & 0.72 +0.92/-0.42 & 1.45 +0.13/-0.16 \\ 
313445 & 0.0118 +0.0054/-0.0032 & 7.11 +0.11/-0.11 & 1.03 +0.94/-0.48 & 0.78 +0.21/-0.20 \\ 
315445 & 0.0001 +0.0008/-0.0001 & 7.22 +0.10/-0.12 & 4.53 +1.56/-1.56 & 0.19 +0.14/-0.12 \\ 
318555 & 0.0121 +0.0065/-0.0039 & 7.25 +0.11/-0.15 & 1.49 +0.74/-0.72 & 0.61 +0.25/-0.32 \\ 
415856 & 0.0037 +0.0027/-0.0026 & 7.17 +0.12/-0.13 & 2.55 +1.15/-0.79 & 0.36 +0.27/-0.22 \\ 
415974 & 0.0012 +0.0020/-0.0012 & 7.12 +0.20/-0.20 & 3.39 +2.02/-1.13 & 0.48 +0.25/-0.27 \\ 
416526 & 0.0060 +0.0024/-0.0014 & 6.91 +0.13/-0.26 & 1.35 +0.99/-0.79 & 0.21 +0.15/-0.16 \\ 
419511 & 0.0000 +0.0000/-0.0000 & 7.38 +0.05/-0.08 & 6.51 +1.00/-1.34 & 0.64 +0.17/-0.27 \\ 
419512 & 0.0000 +0.0006/-0.0000 & 7.77 +0.06/-0.09 & 5.31 +0.99/-1.40 & 0.56 +0.16/-0.24 \\ 
420939 & 0.0001 +0.0011/-0.0001 & 7.70 +0.07/-0.11 & 4.80 +1.31/-1.42 & 0.24 +0.13/-0.15 \\ 
427953 & 0.0100 +0.0033/-0.0027 & 6.96 +0.22/-0.14 & 0.93 +1.09/-0.49 & 0.99 +0.20/-0.18 \\ 
445205 & 0.0000 +0.0003/-0.0000 & 7.62 +0.05/-0.09 & 5.27 +1.21/-1.35 & 0.05 +0.08/-0.04 \\ 
446697 & 0.0000 +0.0001/-0.0000 & 7.37 +0.08/-0.08 & 5.90 +1.28/-1.41 & 0.63 +0.27/-0.43 \\ 
453343 & 0.0061 +0.0020/-0.0013 & 7.00 +0.07/-0.11 & 1.69 +0.77/-0.75 & 0.11 +0.10/-0.08 \\ 
456812 & 0.0025 +0.0025/-0.0023 & 7.54 +0.10/-0.14 & 3.73 +1.07/-1.10 & 0.45 +0.15/-0.27 \\ 
463900 & 0.0030 +0.0006/-0.0007 & 6.95 +0.07/-0.06 & 2.72 +0.63/-0.72 & 0.08 +0.09/-0.05 \\ 
493737 & 0.0076 +0.0019/-0.0029 & 7.31 +0.08/-0.08 & 2.35 +0.81/-0.86 & 0.35 +0.15/-0.15 \\ 
514992 & 0.0013 +0.0009/-0.0011 & 6.80 +0.13/-0.14 & 2.85 +1.35/-1.12 & 0.78 +0.17/-0.19 \\ 
520754 & 0.0000 +0.0002/-0.0000 & 7.69 +0.05/-0.06 & 5.69 +0.88/-1.03 & 0.57 +0.11/-0.11 \\ 
527971 & 0.0105 +0.0054/-0.0033 & 7.21 +0.09/-0.18 & 1.58 +0.90/-0.80 & 0.65 +0.20/-0.25 \\ 
530989 & 0.0003 +0.0014/-0.0003 & 7.58 +0.10/-0.12 & 4.97 +1.43/-1.44 & 0.43 +0.28/-0.26 \\ 
531703 & 0.0103 +0.0041/-0.0041 & 7.28 +0.09/-0.10 & 1.87 +0.82/-0.76 & 0.72 +0.16/-0.20 \\ 
533870 & 0.0004 +0.0013/-0.0004 & 7.01 +0.14/-0.16 & 3.80 +1.95/-1.48 & 0.60 +0.17/-0.22 \\ 
541593 & 0.0014 +0.0051/-0.0014 & 7.66 +0.13/-0.20 & 4.18 +1.69/-1.86 & 0.51 +0.11/-0.17 \\ 
545495 & 0.0020 +0.0007/-0.0010 & 6.76 +0.19/-0.11 & 2.45 +1.06/-0.89 & 0.81 +0.24/-0.19 \\ 
547212 & 0.0003 +0.0021/-0.0003 & 7.33 +0.10/-0.16 & 3.87 +1.60/-1.41 & 0.47 +0.26/-0.32 \\ 
    \end{tabular}
    \label{tab:BAGPIPES}
\end{table*}

\begin{table*}
    \caption{Measured flux densities and errors from the photometric catalogue described in Section \ref{sec:data} for each LSB in the ``Small Kron'' aperture in the NIRCam wide filters F115W, F150W, F200W, F277W, F356W, and F444W in nJy.}
    \centering
    \begin{tabular}{c|c|c|c|c|c|c}
         & \multicolumn{6}{c}{catalogue ``Small Kron'' Aperture $F_{\nu}$ [nJy]} \\ \hline
        ID & F115W & F150W & F200W & F277W & F356W & F444W \\ \hline 
5281 & 3.80 $\pm$ 0.60 & 3.86 $\pm$ 0.48 & 4.80 $\pm$ 0.46 & 4.62 $\pm$ 0.39 & 3.58 $\pm$ 0.40 & 2.35 $\pm$ 0.47\\ 
29405 & 8.35 $\pm$ 0.79 & 7.61 $\pm$ 0.75 & 7.03 $\pm$ 0.71 & 6.86 $\pm$ 0.60 & 5.50 $\pm$ 0.58 & 3.88 $\pm$ 0.62\\ 
34150 & 3.07 $\pm$ 0.74 & 4.24 $\pm$ 0.59 & 5.55 $\pm$ 0.57 & 5.39 $\pm$ 0.40 & 4.51 $\pm$ 0.45 & 3.16 $\pm$ 0.47\\ 
69546 & 13.21 $\pm$ 0.93 & 17.20 $\pm$ 1.07 & 19.96 $\pm$ 0.95 & 24.15 $\pm$ 0.90 & 18.47 $\pm$ 0.79 & 12.46 $\pm$ 0.84\\ 
109405 & 20.47 $\pm$ 1.48 & 29.62 $\pm$ 1.59 & 31.45 $\pm$ 1.65 & 29.24 $\pm$ 0.88 & 24.21 $\pm$ 0.87 & 16.42 $\pm$ 0.92\\ 
113557 & 5.26 $\pm$ 0.47 & 5.28 $\pm$ 0.45 & 5.73 $\pm$ 0.42 & 4.88 $\pm$ 0.30 & 4.62 $\pm$ 0.30 & 3.37 $\pm$ 0.39\\ 
118072 & 14.53 $\pm$ 0.53 & 14.61 $\pm$ 0.61 & 15.00 $\pm$ 0.64 & 12.29 $\pm$ 0.40 & 8.75 $\pm$ 0.40 & 6.13 $\pm$ 0.49\\ 
119192 & 9.56 $\pm$ 0.46 & 11.55 $\pm$ 0.51 & 11.96 $\pm$ 0.53 & 12.34 $\pm$ 0.35 & 8.58 $\pm$ 0.41 & 7.49 $\pm$ 0.43\\ 
127231 & 1.96 $\pm$ 0.27 & 1.41 $\pm$ 0.27 & 3.05 $\pm$ 0.28 & 2.65 $\pm$ 0.22 & 2.20 $\pm$ 0.22 & 2.17 $\pm$ 0.28\\ 
143098 & 3.82 $\pm$ 0.47 & 5.60 $\pm$ 0.52 & 5.23 $\pm$ 0.53 & 5.55 $\pm$ 0.43 & 3.91 $\pm$ 0.43 & 3.41 $\pm$ 0.55\\ 
160858 & 21.73 $\pm$ 1.44 & 26.77 $\pm$ 1.30 & 27.61 $\pm$ 1.31 & 30.92 $\pm$ 1.16 & 22.70 $\pm$ 1.02 & 17.44 $\pm$ 1.06\\ 
164373 & 13.80 $\pm$ 0.63 & 12.37 $\pm$ 0.59 & 11.86 $\pm$ 0.68 & 13.14 $\pm$ 0.58 & 10.34 $\pm$ 0.56 & 7.76 $\pm$ 0.63\\ 
171611 & 7.11 $\pm$ 0.86 & 8.02 $\pm$ 0.80 & 9.31 $\pm$ 0.89 & 9.79 $\pm$ 0.67 & 9.44 $\pm$ 0.62 & 7.68 $\pm$ 0.66\\ 
175180 & 13.41 $\pm$ 0.58 & 12.27 $\pm$ 0.53 & 11.64 $\pm$ 0.52 & 11.57 $\pm$ 0.55 & 7.15 $\pm$ 0.51 & 4.62 $\pm$ 0.56\\ 
183481 & 8.03 $\pm$ 0.53 & 7.58 $\pm$ 0.50 & 8.93 $\pm$ 0.56 & 10.04 $\pm$ 0.56 & 8.03 $\pm$ 0.52 & 5.26 $\pm$ 0.58\\ 
189735 & 13.10 $\pm$ 1.10 & 15.90 $\pm$ 1.21 & 15.28 $\pm$ 1.34 & 15.75 $\pm$ 1.02 & 10.92 $\pm$ 0.93 & 7.27 $\pm$ 0.88\\ 
190476 & 21.77 $\pm$ 1.20 & 26.40 $\pm$ 1.21 & 25.37 $\pm$ 1.16 & 27.23 $\pm$ 0.98 & 20.39 $\pm$ 0.86 & 16.98 $\pm$ 0.87\\ 
196612 & 7.42 $\pm$ 0.45 & 8.96 $\pm$ 0.44 & 10.24 $\pm$ 0.46 & 10.16 $\pm$ 0.36 & 7.85 $\pm$ 0.37 & 6.76 $\pm$ 0.48\\ 
198340 & 9.51 $\pm$ 0.48 & 7.25 $\pm$ 0.57 & 7.44 $\pm$ 0.46 & 7.48 $\pm$ 0.47 & 4.57 $\pm$ 0.45 & 3.87 $\pm$ 0.72\\ 
198413 & 8.84 $\pm$ 0.84 & 12.05 $\pm$ 1.42 & 12.01 $\pm$ 0.92 & 13.51 $\pm$ 0.83 & 8.46 $\pm$ 0.67 & 5.14 $\pm$ 0.88\\ 
199235 & 11.25 $\pm$ 0.92 & 14.23 $\pm$ 0.98 & 14.20 $\pm$ 0.95 & 15.04 $\pm$ 0.58 & 10.90 $\pm$ 0.57 & 7.38 $\pm$ 0.64\\ 
207083 & 12.65 $\pm$ 0.66 & 12.78 $\pm$ 0.66 & 13.49 $\pm$ 0.64 & 13.05 $\pm$ 0.39 & 9.80 $\pm$ 0.42 & 5.54 $\pm$ 0.49\\ 
207340 & 36.17 $\pm$ 1.09 & 34.95 $\pm$ 1.10 & 32.07 $\pm$ 1.11 & 31.90 $\pm$ 1.17 & 21.66 $\pm$ 1.11 & 19.60 $\pm$ 1.11\\ 
208568 & 4.81 $\pm$ 0.42 & 5.30 $\pm$ 0.42 & 4.93 $\pm$ 0.43 & 4.41 $\pm$ 0.33 & 3.95 $\pm$ 0.33 & 1.91 $\pm$ 0.40\\ 
209020 & 20.24 $\pm$ 1.20 & 22.32 $\pm$ 1.33 & 23.43 $\pm$ 1.22 & 22.18 $\pm$ 1.03 & 14.87 $\pm$ 1.09 & 8.76 $\pm$ 0.99\\ 
210372 & 18.08 $\pm$ 1.12 & 21.34 $\pm$ 1.13 & 20.99 $\pm$ 1.17 & 21.08 $\pm$ 0.82 & 15.49 $\pm$ 0.77 & 10.01 $\pm$ 0.83\\ 
216996 & 9.52 $\pm$ 0.60 & 10.16 $\pm$ 0.61 & 9.74 $\pm$ 0.73 & 9.57 $\pm$ 0.55 & 7.17 $\pm$ 0.55 & 5.24 $\pm$ 0.63\\ 
217509 & 10.27 $\pm$ 0.67 & 9.82 $\pm$ 0.69 & 11.96 $\pm$ 0.85 & 14.22 $\pm$ 0.66 & 12.58 $\pm$ 0.60 & 11.06 $\pm$ 0.74\\ 
217847 & 28.61 $\pm$ 1.02 & 35.73 $\pm$ 1.17 & 34.19 $\pm$ 1.14 & 37.78 $\pm$ 0.86 & 25.92 $\pm$ 0.76 & 18.84 $\pm$ 0.84\\ 
235455 & 42.38 $\pm$ 2.34 & 47.62 $\pm$ 2.57 & 54.41 $\pm$ 2.77 & 55.35 $\pm$ 2.00 & 37.72 $\pm$ 1.92 & 25.28 $\pm$ 1.94\\ 
285675 & 3.50 $\pm$ 0.49 & 6.89 $\pm$ 0.53 & 7.47 $\pm$ 0.54 & 10.90 $\pm$ 0.36 & 8.97 $\pm$ 0.36 & 7.05 $\pm$ 0.45\\ 
301713 & 7.47 $\pm$ 0.70 & 12.57 $\pm$ 0.78 & 12.91 $\pm$ 0.83 & 13.97 $\pm$ 0.98 & 12.00 $\pm$ 0.85 & 14.73 $\pm$ 1.16\\ 
313445 & 11.91 $\pm$ 1.03 & 9.30 $\pm$ 0.99 & 11.00 $\pm$ 1.02 & 16.47 $\pm$ 0.71 & 13.52 $\pm$ 0.69 & 8.97 $\pm$ 0.88\\ 
315445 & 12.50 $\pm$ 1.40 & 13.98 $\pm$ 1.19 & 13.38 $\pm$ 1.32 & 13.06 $\pm$ 0.86 & 10.45 $\pm$ 0.90 & 6.72 $\pm$ 1.07\\ 
318555 & 5.79 $\pm$ 0.31 & 6.45 $\pm$ 0.48 & 7.36 $\pm$ 0.55 & 6.41 $\pm$ 0.34 & 6.38 $\pm$ 0.33 & 4.55 $\pm$ 0.48\\ 
415856 & 7.67 $\pm$ 0.61 & 7.62 $\pm$ 0.44 & 8.66 $\pm$ 0.55 & 8.77 $\pm$ 0.39 & 6.01 $\pm$ 0.35 & 4.19 $\pm$ 0.49\\ 
415974 & 13.52 $\pm$ 2.72 & 15.69 $\pm$ 1.00 & 15.74 $\pm$ 1.10 & 15.79 $\pm$ 0.79 & 10.87 $\pm$ 0.75 & 9.20 $\pm$ 0.79\\ 
416526 & 9.57 $\pm$ 0.67 & 10.61 $\pm$ 0.62 & 9.74 $\pm$ 0.91 & 9.00 $\pm$ 0.46 & 6.55 $\pm$ 0.39 & 4.58 $\pm$ 0.80\\ 
419511 & 14.08 $\pm$ 0.99 & 18.48 $\pm$ 1.01 & 17.98 $\pm$ 0.97 & 17.15 $\pm$ 0.64 & 11.65 $\pm$ 0.70 & 8.94 $\pm$ 0.76\\ 
419512 & 19.57 $\pm$ 1.47 & 23.21 $\pm$ 1.23 & 23.33 $\pm$ 2.01 & 26.74 $\pm$ 0.89 & 19.44 $\pm$ 0.81 & 14.20 $\pm$ 0.94\\ 
420939 & 24.78 $\pm$ 1.17 & 24.86 $\pm$ 1.39 & 29.27 $\pm$ 1.94 & 28.53 $\pm$ 1.05 & 18.76 $\pm$ 0.92 & 13.79 $\pm$ 1.08\\ 
427953 & 10.06 $\pm$ 2.07 & 13.98 $\pm$ 1.22 & 15.39 $\pm$ 1.46 & 17.94 $\pm$ 0.83 & 13.90 $\pm$ 0.70 & 13.99 $\pm$ 0.82\\ 
445205 & 15.84 $\pm$ 0.84 & 18.14 $\pm$ 0.86 & 21.88 $\pm$ 0.91 & 21.26 $\pm$ 0.57 & 13.33 $\pm$ 0.57 & 7.76 $\pm$ 0.62\\ 
446697 & 9.24 $\pm$ 0.70 & 10.82 $\pm$ 0.76 & 12.36 $\pm$ 0.78 & 12.24 $\pm$ 0.57 & 7.92 $\pm$ 0.53 & 6.33 $\pm$ 0.66\\ 
453343 & 9.86 $\pm$ 0.89 & 8.88 $\pm$ 0.79 & 8.70 $\pm$ 0.76 & 8.15 $\pm$ 0.60 & 5.92 $\pm$ 0.59 & 3.35 $\pm$ 0.65\\ 
456812 & 23.86 $\pm$ 1.20 & 28.71 $\pm$ 1.29 & 26.41 $\pm$ 1.34 & 31.73 $\pm$ 1.13 & 21.41 $\pm$ 0.98 & 17.65 $\pm$ 1.12\\ 
463900 & 9.16 $\pm$ 0.43 & 11.00 $\pm$ 0.42 & 11.02 $\pm$ 0.44 & 10.52 $\pm$ 0.32 & 7.27 $\pm$ 0.32 & 5.41 $\pm$ 0.40\\ 
493737 & 29.07 $\pm$ 2.42 & 31.70 $\pm$ 2.43 & 37.20 $\pm$ 2.04 & 37.62 $\pm$ 1.89 & 28.33 $\pm$ 2.11 & 20.05 $\pm$ 2.36\\ 
514992 & 4.10 $\pm$ 0.58 & 4.60 $\pm$ 0.41 & 4.46 $\pm$ 0.45 & 4.46 $\pm$ 0.32 & 3.35 $\pm$ 0.31 & 3.28 $\pm$ 0.36\\ 
520754 & 19.45 $\pm$ 1.24 & 27.96 $\pm$ 1.35 & 23.40 $\pm$ 1.39 & 24.12 $\pm$ 1.02 & 21.22 $\pm$ 0.93 & 12.63 $\pm$ 1.02\\ 
527971 & 9.49 $\pm$ 0.80 & 10.13 $\pm$ 0.82 & 9.03 $\pm$ 0.83 & 11.43 $\pm$ 0.59 & 8.11 $\pm$ 0.55 & 6.78 $\pm$ 0.69\\ 
530989 & 27.32 $\pm$ 1.31 & 35.70 $\pm$ 1.33 & 38.56 $\pm$ 1.30 & 36.76 $\pm$ 1.14 & 25.24 $\pm$ 1.00 & 21.20 $\pm$ 1.14\\ 
531703 & 13.51 $\pm$ 0.84 & 13.75 $\pm$ 0.84 & 12.71 $\pm$ 0.84 & 13.94 $\pm$ 0.73 & 10.57 $\pm$ 0.65 & 8.52 $\pm$ 0.79\\ 
533870 & 3.56 $\pm$ 0.43 & 4.71 $\pm$ 0.39 & 4.42 $\pm$ 0.41 & 4.60 $\pm$ 0.30 & 3.83 $\pm$ 0.39 & 2.81 $\pm$ 0.37\\ 
541593 & 24.78 $\pm$ 1.40 & 30.35 $\pm$ 1.41 & 32.21 $\pm$ 1.53 & 29.34 $\pm$ 0.88 & 21.39 $\pm$ 0.85 & 18.04 $\pm$ 0.88\\ 
545495 & 4.31 $\pm$ 0.47 & 5.42 $\pm$ 0.44 & 6.84 $\pm$ 0.42 & 6.77 $\pm$ 0.34 & 5.56 $\pm$ 0.36 & 4.71 $\pm$ 0.43\\ 
547212 & 7.70 $\pm$ 0.64 & 7.98 $\pm$ 0.67 & 9.72 $\pm$ 0.69 & 10.96 $\pm$ 0.60 & 7.46 $\pm$ 0.54 & 5.09 $\pm$ 0.70\\
    \end{tabular}
    \label{tab:fluxes}
\end{table*}

%%%%%%%%%%%%%%%%%%%%%%%%%%%%%%%%%%%%%%%%%%%%%%%%%%

% Don't change these lines
\bsp	% typesetting comment
\label{lastpage}
\end{document}